\documentclass[apj,numberedappendix]{emulateapj}
                                                                
\usepackage[utf8]{inputenc}
\usepackage{graphicx}
\usepackage{amsmath}
\usepackage{amssymb}

\usepackage[usenames,dvipsnames]{xcolor}
\newcommand{\rrbf}[1]{#1}

\def\mfmtk{{\sc Morfometryka}}
\def\mfm{{\sc mfmtk}}

\begin{document}

\shorttitle{\textsc{Morfometryka}}
\shortauthors{Ferrari et al.}

\title{Morfometryka -- A New Way of Establishing Morphological Classification of Galaxies}

\author{F. Ferrari}
\affil{IMEF -- FURG, Rio Grande, RS, Brazil}
\email{fabricio@ferrari.pro.br}

\author{R. R. de Carvalho}
\affil{INPE/MCT, Divis\~ao de Astrof\'isica, S. J. dos Campos, Brazil}

\and 

\author{M. Trevisan}
\affil{INPE/MCT, Divis\~ao de Astrof\'isica, S. J. dos Campos, Brazil}
\affil{Institut d’Astrophysique de Paris, F-75014, Paris, France}

\begin{abstract}

We present an extended morphometric system to automatically classify
galaxies from astronomical images.  The new system includes the
original and modified versions of the CASGM coefficients
(Concentration $C_1$, Asymmetry $A_3$, and Smoothness $S_3$), and the
new parameters entropy, $H$, and spirality $\sigma_\psi$.  The new
parameters $A_3$, $S_3$ and $H$ are better to discriminate galaxy
classes than $A_1$, $S_1$ and $G$, respectively. The new parameter
$\sigma_\psi$ captures the amount of non-radial pattern on the image
and is almost linearly dependent on T-type.  Using a sample of spiral
and elliptical galaxies from the Galaxy Zoo project as a training set,
we employed the Linear Discriminant Analysis (LDA) technique to
classify \citet[4478 galaxies]{EFIGI}, \citet[14123 galaxies]{NA}
and SDSS Legacy (779,235 galaxies) samples. The cross-validation test
shows that we can achieve an accuracy of more than 90\% with our
classification scheme. Therefore, we are able to define a plane in the
morphometric parameter space that separates the elliptical and spiral
classes with a mismatch between classes smaller than 10\%.  We use the
distance to this plane as a morphometric index (M$_{\rm i}$) and we
show that it follows the human based T-type index very closely. We
calculate morphometric index M$_{\rm i}$ for $\sim$780k galaxies from
SDSS Legacy Survey - DR7. We discuss how M$_{\rm i}$ correlates with
stellar population parameters obtained using the spectra available
from SDSS-DR7.
\end{abstract}

\keywords{galaxies: morphology,  morphometry, classification;   methods: statistical}

\section{Introduction}
\label{sec:intro}

The study of the formation and evolution of galaxies in general
requires their systematic observations over a large redshift domain.
Datasets for local (today$'$s systems) and distant galaxies (their
actual progenitors) must be consistently gathered to avoid biases, a
procedure that requires a knowledge of the very evolution we are
seeking to understand. From an astrophysical perspective, mechanisms
regulating star formation, e.g. ram-pressure \citep{GunnGott},
harassment \citep{moore}, starvation \citep{starvation}, depend on the
distance from the center of the potential well of a cluster; their
effects on the stellar population of a galaxy depend on how
efficiently the local environment is capable of removing the
interstellar gas and affect the star formation history. Thus,
morphology, in a general sense, is just a snapshot reflecting all
these processes imprinted in the galaxy image at a given time.

Traditionally, galaxy morphology has been addressed visually: an
expert examines the images of galaxies and identifies features (or
absence of them, in the case of early-type galaxies) which distinguish
the object as belonging to a specific class, as done in \cite{hubble};
\cite{devauc1959}; \cite{sandage}; \cite{vandenbergh};
\cite{zoo1,zoo2}, among many others. This classification paradigm is
strongly subjective, prone to errors and cannot be applied to the
number of galaxies present in modern surveys For instance,
\cite{FASANO} compare their classification with that of \citet[][RC3
  catalog]{RC3} (see their Figure 2). As it is clearly seen there is
an uncertainty of approximately 1 in T-type within \cite{FASANO}
authors and about 2.5 between RC3 and \cite{FASANO}.  Comparison
between EFIGI and NA2010, for 1438 galaxies in common \citep[Figure
  32]{EFIGI}, exhibits a similar sort of inconsistency, with an
uncertainty between 2 and 3 in T-type.  This tell us that, for
example, visual classification does not agree when distinguishing an
S0 from an Sa or an E from an S0. Thus, it is imperative to quantify
the morphology of a galaxy as a measurable quantity -- morphometry --
that can be coded in an algorithm. \rrbf{However, in spite of its
  uncertainties, visual classification is still important, because
  automated techniques would have difficulty doing classifications
  like ${\rm (R_1R_2')SAB(r,nr)0/a}$ which is much more valuable than
  just knowing whether a galaxy is S, S0, or E.  Also, regarding RC3,
  it should be noted that its classifications were made based on
  photographic plates or sky survey charts. Modern digital images
  allow greater consistency between multiple classifiers and have the
  potential to greatly improve on RC3.}

Two approaches for galaxy morphometry have been widely explored
recently: parametric -- those which model the light distribution as a
bulge plus disk plus other less important components representing a
few percent of the total light \citep[e.g.][]{galfit,gim2d};
non-parametric -- those which use the measured properties of the light
distribution, like concentration, asymmetry
\citep[e.g.][]{abraham1996}. Each approach has its virtues and vices,
as discussed for example in \cite{Andrae}.

One relatively successful non-parametric system is the concentration,
asymmetry, smoothness, Gini and M20 (CASGM) system, presented in
\citet{abraham1994,abraham1996}, \cite{conselice2000} and
\citet{Lotz2004}.  This basic set has been enlarged with other
quantities such as the S\'ersic model parameters \citep{Sersic}, and
the Petrosian radius \citep{petrosian}, among others. These quantities
may not work properly in the high redshift regime and this has been
studied in recent papers \cite[e.g.][]{freeman}. These authors use
Multi-mode, Intensity and Deviation statistics, MID, to detect
disturbances in the galaxy light distribution and show that it is very
effective at z $\sim$2. 

The previously mentioned way of establishing galaxy morphology answers
two immediate needs. First, it is possible to reproduce human
classification by positioning the galaxies in the space of these
parameters. In such a supervised classification, a set of visually
classified galaxies are used to train a discriminant function that
will assign to each new galaxy a probability of belonging to each
class. The second reason for establishing a galaxy morphometry system
is that we can seek structures, in the quantitative morphology
parameter space, that may yield clues for the physical reasons for
their formation and evolution that are not visible in the currently
human-based mode. Further, a system such as the Hubble tuning
fork classification does not account for all the details that we can
currently measure in galaxy images, and it does not hold as we go deep
even at a moderate redshift $z=0.25$. \rrbf{ To evaluate this, a new
  quantitative classification }procedure is needed, both to handle the
large amount of data becoming available with the new surveys, and also
to help us find the physical processes driving galaxy evolution.

The paper is organized as follows: in Section \ref{sec:related} we
discuss similar works; in Section \ref{sec:data} we describe the
datasets used; in Section \ref{sec:quantitative.morpho} we define new
nonparametric methods to quantify galaxy morphology; in Section
\ref{sec:morfometryka} we present the \mfmtk\ algorithm. We apply the
\mfmtk\ code to galaxy samples described in Section
\ref{sec:classification}, where we also test the robustness of the
measured parameters and explore the ability of them to classify galaxy
morphologies. In Section \ref{sec:Mi} we propose a new Morphometric
Index M$_{\rm i}$. In Section \ref{sec:results} we compare M$_{\rm i}$
with other physical parameters and summary is presented in Section
\ref{sec:summary}.

\section{Related Work}
\label{sec:related}

There have been several attempts to classify galaxies automatically,
beginning with \citet{abraham1994,abraham1996}, among others. Here we
briefly mention recent works based on machine learning and on
morphometric parameters. The list is not meant to be exhaustive but
rather to present different approaches followed in the last few years.

\cite{huertas} used a system based on colors, shapes and concentration
to train a support vector machine to classify $\sim$700k galaxies from
the SDSS DR7 spectroscopic sample. For each galaxy, they estimate the
probabilities of being E, S0, Sab or Scd. It is not a pure
morphometric classification, since it includes colors.

\cite{Scarlata} analyses 56,000 COSMOS galaxies with the ZEST
algorithm, using five nonparametric diagnostics ($A$, $C_1$, $G$,
$M_{20}$, $q$) and S\'ersic index $n$. They perform
Principal Component Analysis (PCA) and classify galaxies with 3
principal components. They find contamination between galaxy classes
in parameters space (see their Fig.10), although they do not state
clearly the success rate of the classification.

\cite{Andrae} present a detailed analysis of several critical issues
when dealing with galaxy morphology and classification. Several
morphological features are intertwined and cannot be estimated
independently.  They show the dependence between $C$ and $n$, which is
also presented here in a different form in Appendix \ref{app:C_and_n}.
The authors claim that parameter based approaches are better for
classification, and state that a system such as CASGM has serious
problems. However, they do not show it in practice.

\cite{Dieleman} present a Neural Network machine to reproduce Galaxy
Zoo classification. They work directly in pixel space, using a
rotation invariant convolution that minimizes sensitiveness to changes
in scale, rotation, translation and sampling of the image. The
algorithm obtains an accuracy of 99\% relative to the Galaxy Zoo human
classification; however, since the human classification is also error
prone, as discussed in Section \ref{sec:intro}, their algorithm
reproduces also the errors in the human classification.

\cite{freeman} introduced MID (multimode, intensity and deviation)
statistics designed to detect disturbed morphologies, and then
classify 1639 galaxies observed with HST WFPC3 with a random
forest. It is one of the few works that state the detailed classifier
performance, in terms of the confusion matrix coefficients.

\section{Data and Sample Selection}
\label{sec:data}

\begin{table}
	\begin{tabular}{|p{2cm}|r|r|r|r|}
		\hline             & EFIGI       & NA      & LEGACY    & LEGACY$-zr$ \\ \hline  
		Total                 & 4\,458   & 14\,034 & 804\,974  & 337\,097 \\ 
		\textsc{\mfm}        & 4\,214   & 12\,729 & 779\,235  & 327\,937 \\ 
		\textsc{\mfm}+Zoo & 1\,856   &  8\,792 & 245\,206  & 125\,417 \\ \hline
	\end{tabular} 
	\label{tab:datasets}
	\caption{Number of objects in the databases used in this
          work. MFMTK are objects for which \mfmtk\ made measurements;
          Zoo means objects classified as E or S by Galaxy Zoo. }
\end{table}

We use several databases derived from SDSS DR7 \citep{SDSS7}, for
which we analyze $r$ band images. They are: the \citet{EFIGI} database
(hereafter EFIGI); the \cite{NA} database (hereafter NA); and the SDSS
DR7 complete Legacy database and a volume limited subsample, hereafter
referred as LEGACY and LEGACY--$zr$, respectively. We also use the
Galaxy Zoo collaborative project visual classification
\citep{zoo1,zoo2}. The number of galaxies in the original databases,
those succesfully processed by \mfmtk\ and those that have a Galaxy
Zoo classification are listed in Table~\ref{tab:datasets}. 
 
The databases are used with different purposes, namely, training,
validation and classification. In training phase, the galaxies in the
databases that have Galaxy Zoo classification are used to train a
classifier machine (Section \ref{sec:LDA}). For validation, we use a
cross validation scheme to attest how well our classifier performed
compared to Galaxy Zoo human classification. In the classification
stage, we use the trained classifier to linearly separate LEGACY
galaxies in two classes (elliptical--E or spiral--S) in the
morphometric parameters space (Section \ref{sec:LDA}). The galaxy
distance to the separating hyperplane is then proposed as a
morphometric index $M_{\rm i}$ (Section \ref{sec:Mi}).  The databases
EFIGI and NA, for which we have T-type values, are further used to
support out argument that $M_{\rm i}$, based on the classifier
discriminant function, can reflect the galaxy morphological type.

The classification scheme from the Galaxy Zoo project
\citep{zoo1,zoo2} was used to train our supervised morphometric
classifier.  The Galaxy Zoo project provides simple morphological
classifications of nearly 900,000 galaxies drawn from the SDSS-DR6.
  
Below we discuss each sample in detail.

\subsection{The EFIGI sample}
 
The EFIGI catalog was specifically designed to sample all Hubble
morphological types. It provides detailed morphological information of
galaxies selected from standard surveys and catalogues (Principal
Galaxy Catalogue, Sloan Digital Sky Survey, Value-Added Galaxy
Catalogue, HyperLeda, and the NASA Extragalactic Database). The sample
is essentially limited in apparent diameter, and offers a detailed
view of the whole Hubble sequence. The final EFIGI sample comprises
4458 galaxies for which there is imaging in all the $ugriz$ bands in
the SDSS-DR4 database. For these galaxies, the EFIGI reference dataset
provides visually estimated morphological information as well as
re-sampled SDSS imaging data. The photometric catalog is more than ~
80\% complete for galaxies with $10 < m_{{\rm petro},g} < 14$, where
$m_{{\rm petro},g}$ is the Petrosian magnitude in the $g$ band.
 
\subsection{The NA sample}

\cite{NA} provide detailed visual classifications for 14034 galaxies
selected from the SDSS spectroscopic main sample described in Strauss
et al. (2002). They used the SDSS-DR4 photometry catalogs to select
all spectroscopically targeted galaxies in the redshift range $0.01 <
z < 0.1$ down to an apparent extinction-corrected magnitude limit of
$g < 16$ mag. Objects mistakenly classified as galaxies have been
removed, leading to the final sample of 14,034 galaxies. Their final
catalog provides T-Types, the existence of bars, rings, lenses, tails,
warps, dust lanes, arm flocculence, and multiplicity for all galaxies.
 
\subsection{The SDSS LEGACY and LEGACY--$zr$ samples}
 
Our target sample of galaxies was retrieved from SDSS-DR7
\citep{SDSS7} by selecting all objects spectroscopically classified as
galaxies (see Appendix \ref{sec:stamp} for full query).  SDSS Frames
and psFields were obtained and stamps and PSF (point spread function)
generated from them (see Appendix \ref{app:mfmtk} for details). Our
final catalog comprises 804,974 objects.
 
The subsample LEGACY--$zr$ is volume limited at redshift $z<0.1$ and
$m_{{\rm petro},r} < 17.78$, where $m_{{\rm petro},r}$ is the
extinction corrected Petrosian magnitude in the $r$ band.  This
magnitute limit roughly corresponds to the magnitude at which the SDSS
spectroscopy is complete \citep{SDSS7spec}.  The redshift limit of $z
< 0.1$ provides a complete sample for $M_{\rm petro,r} < -20.5$, where
$M_{\rm petro,r}$ is the SDSS Petrosian absolute magnitude in the $r$
band.
 
For 570,685 galaxies, those for which \texttt{zWarning=0} in the SDSS
database, we derived ages, metallicities, stellar masses and velocity
dispersions using the spectral fitting code \textsc{starlight}
\citep{Cid2005}.  Before running the code, the observed spectra are
corrected for foreground extinction and de-redshifted, and the single
stellar population (SSP) models are degraded to match the
wavelength-dependent resolution of the SDSS spectra, as described in
\citet{labarbera}. We adopted the \citet{Cardelli.etal:1989}
extinction law, assuming $R_{\rm V} = 3.1$.
 
We used SSP models based on the Medium resolution INT Library of
Empirical Spectra \citep[MILES, ][]{SanchezBlazquez.etal:2006}, using
the code presented in \citet{Vazdekis.etal:2010}, using version 9.1
\citep{FalconBarroso.etal:2011}.  They have a spectral resolution of
$\sim 2.5$ \AA, almost constant with wavelength.  We selected models
computed with \citet{Kroupa:2001} Universal IMF with slope $= 1.30$,
and isochrones by \citet{Girardi.etal:2000}.  The basis grids cover
ages in the range of $0.07 - 14.2$~Gyr, with constant $\log({\rm
  Age})$ steps of 0.2.  We selected SSPs with metallicities
[M/H]~=$\{-1.71, -0.71, -0.38, 0.00, +0.20\}$.

\section{Quantitative galaxy morphology}
\label{sec:quantitative.morpho}

The basic morphometric measurements of the CASGM system are fully
described by \citet{abraham1994,abraham1996}, \citet{bershady},
\citet{conselice2000} and \cite{Lotz2004}, among others.  Relevant
modifications of these parameters and the new parameters introduced in
this work are discussed in this section.

We define the region with the same axis ratio and position angle as
the galaxy (see Section \ref{sec:stamp}) and with major axis equal
$N_{\!R_p} R_p$, where $R_p$ is the Petrosian Radius and
$N_{\!R_p}=2$, as the \textbf{Petrosian region}. Most measurements are
made with pixels in this regions, except if otherwise stated. A central
region of the size of the PSF FWHM is masked before calculating $A$,
$S$ and $\sigma_\psi$.

\subsection{Concentration $C_1$ and $C_2$} 

The concentration index $C$ is the ratio of the circular radii
containing two fractions of the total flux of the galaxy \citep{kent},
where these percentages are chosen to maximize the distinction between
systems and minimize seeing effects. The concentration depends on the
determination of the radius that contain some fraction of some measure
of the total luminosity of the galaxy. In this work, we have adopted
the Petrosian luminosity as the total luminosity $L_T$, which is the
maximum value of $L(R)$ inside the Petrosian region. The measured
$L(R)$ is spline interpolated and then the point where it attains some
fraction $f$ of $L_T$ is found by evaluating the spline at the point. 
In this way we obtain $R_{20}$, $R_{50}$, $R_{80}$ and
$R_{90}$ and finally
\[
C_1 =  \log_{10}{\left(\frac{R_{80}}{R_{20}}\right)} \qquad \text{and} \qquad 
C_2 =  \log_{10}{\left(\frac{R_{90}}{R_{50}}\right)}.
\]  
Note that we dropped the factor 5 usually in the definition of $C$, so
that all morphometric measurements used will fall approximately in the range $[0,1]$, and
thus statistic standardization would have little effect and may be
optional.  The concentration $C_1$ is more sensitive to seeing effect
that is more pronounced in the central regions and thus on $R_{20}$;
$C_2$ is more sensitive to noise that is more important in the outer
regions and thus on the measure of $R_{90}$.

\subsection{Asymmetry $A_1$, $A_2$, $A_3$}

The asymmetry coefficient $A$ is determined comparing a source image
with its rotated counterpart.  We measure the asymmetry $A_1$, as
defined by \cite{abraham1996}, with the exception that we do not
subtract the background asymmetry, for we \rrbf{find} that this
procedure makes the asymmetry estimation unstable and sensitive to the
selected sky \rrbf{(an hence to the stamp)} size. Instead, we consider
only the galaxy portion inside the Petrosian region. Even so, $A_1$
depends heavily on the noise and on the image sampling. To address
this problems we used two new asymmetry measurements defined as
\begin{equation}
A_2 = 1 - r(I,I_{\pi})
\end{equation}
and 
\begin{equation}
A_3 = 1 - s(I,I_{\pi})
\end{equation}
where $r()$ and $s()$ are the Pearson and Spearman correlation
coefficients \citep{NR}, respectively, $I$ is the image and $I_\pi$
its $\pi$-rotated version.  The rationale behind this formulation is
that those pixels made up mostly of noise will not contribute to $A_2$
and $A_3$ since the correlation between them will tend to
zero. Further, correlation coefficients are more immune to convolution
and thus less affected by seeing effects. The Pearson coefficient
tends to accumulate close to unity and so $A_2$ has proven not so
useful as $A_1$ and $A_3$. The center of rotation is chosen to
minimize the asymmetry measurements.

\subsection{Gini coefficient}
The Gini coefficient $G$ measures the flux distribution among the
pixels of a galaxy image.  The Gini coefficient for the image pixels
in the Petrosian region is calculated exactly as shown in
\citet{Lotz2004}, i.e., for $n$ pixels with values $I_i$ in increasing
order we have
\[
G = \frac{1}{n(n-1)\, \bar{I} }\  \sum\limits_{i}^{n} (2i-n-1)\, I_i,
\] 
where $\bar{I}$ is the average value.

\subsection{Smoothness}
The smoothness coefficient $S$ (a.k.a clumpiness) in general measures
the small scale structures in the galaxy image. Here we consider three
different measures of smoothness. $S_1$ is calculated as shown in \cite{Lotz2004}, 
except that the filter used is a Hamming window
\citep{hamming} with size $\lceil R_p/4 \rceil$. Following the same
reasoning as for asymmetry, we define the modified smoothness $S_2$
and $S_3$ as

\begin{equation}
S_2 = 1 - r(I,I^F )
\end{equation}
and 
\begin{equation}
S_3 = 1 - s(I,I^F)
\end{equation}
where $I^F$ is the filtered image. As with asymmetry, $S_3$ has proven
to be more useful than $S_2$.

\subsection{Entropy}
The entropy of information $H$ (Shannon entropy, e.g. \citet{Bishop})
is used here to quantify the distribution of pixel values in the
image. For a random variable $I$, the entropy $H(I)$ is the expected
value of the information $\log[p(I)]$
\begin{equation}
\label{eq:entropy}
H(I) = - \sum_k^K \; p(I_k) \; \log[p(I_k)],
\end{equation}
where $p(I_k)$ is the probability of occurrence of the value $I_k$,
$k$ refers to a specific value and $K$ is the number of bins
considered. For discrete variables, $H$ reaches the maximum value for
a uniform distribution, when $p(I_k) = 1/K$ for all $k$ and hence
$H_{\rm max}=\log K$. The minimum entropy is that of a delta function,
for which $H=0$.  We then have the normalized entropy
\begin{equation}
\label{eq:entropynorm}
\widetilde{H}(I) = \frac{H(I)}{H_{\rm max}} \qquad 0 \leqslant \widetilde{H}(I) \leqslant 1.
\end{equation}
Smooth galaxies will have low $H$ while clumpy will have high $H$.

\subsection{Spirality $\sigma_\psi$}

\begin{figure*}
	\centering
	\includegraphics[width=\linewidth]{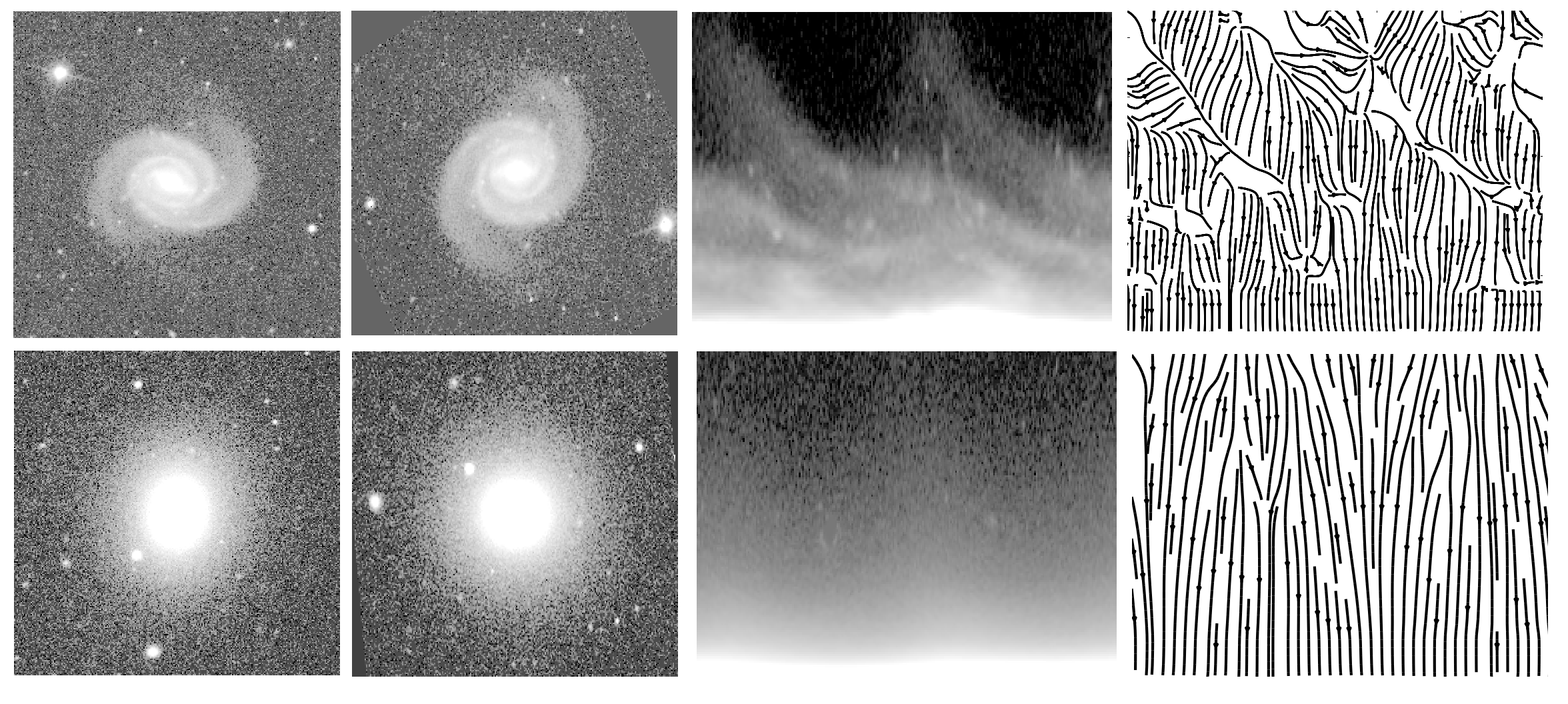}
	\caption{Illustration about how spirality is measured. EFIGI
          spiral SB PGC 2182 above and lenticular PGC 2076 below. From
          left to right: original image, standardized image ($q=1$,
          $PA=0^\circ{}$), image in polar coordinates, gradient field
          of polar image.}
	\label{fig:MontagemES}
\end{figure*}

None of the CASGM parameters take into account the spiral arms, rings
and bars in galaxies, \rrbf{ albeit they are a major and important
  emphasis of human based classification.} We devise a parameter to
take it into account. As done in Shamir (2011), we first transform the
standardized galaxy image to polar coordinates $(r,\theta)$. In
$(r,\theta)$ space a bulge appears as a band in the lower region of
the diagram, a bar as two vertical lines and spiral arms as inclined
bands. See Figure \ref{fig:MontagemES}. We then calculate the gradient
magnitude $|\nabla I|$ and direction $\vec{\psi}$ fields of the polar
image.  Most points in this direction field $\vec{\psi}$ for an
elliptical galaxy will point to the bottom, whilst for a spiral galaxy
there will be many orientations corresponding to arms, rings and
bars. The standard deviation $\sigma_\psi$ for the field direction
values will be smaller for an elliptical compared to that of a spiral,
and hence can be used to estimate the amount of characteristic
structures. To avoid regions of noise, we make the measurements in
regions where the gradient magnitude is greater than the median of the
magnitude field.

\begin{figure}
	\centering
	\includegraphics[width=0.50\textwidth]{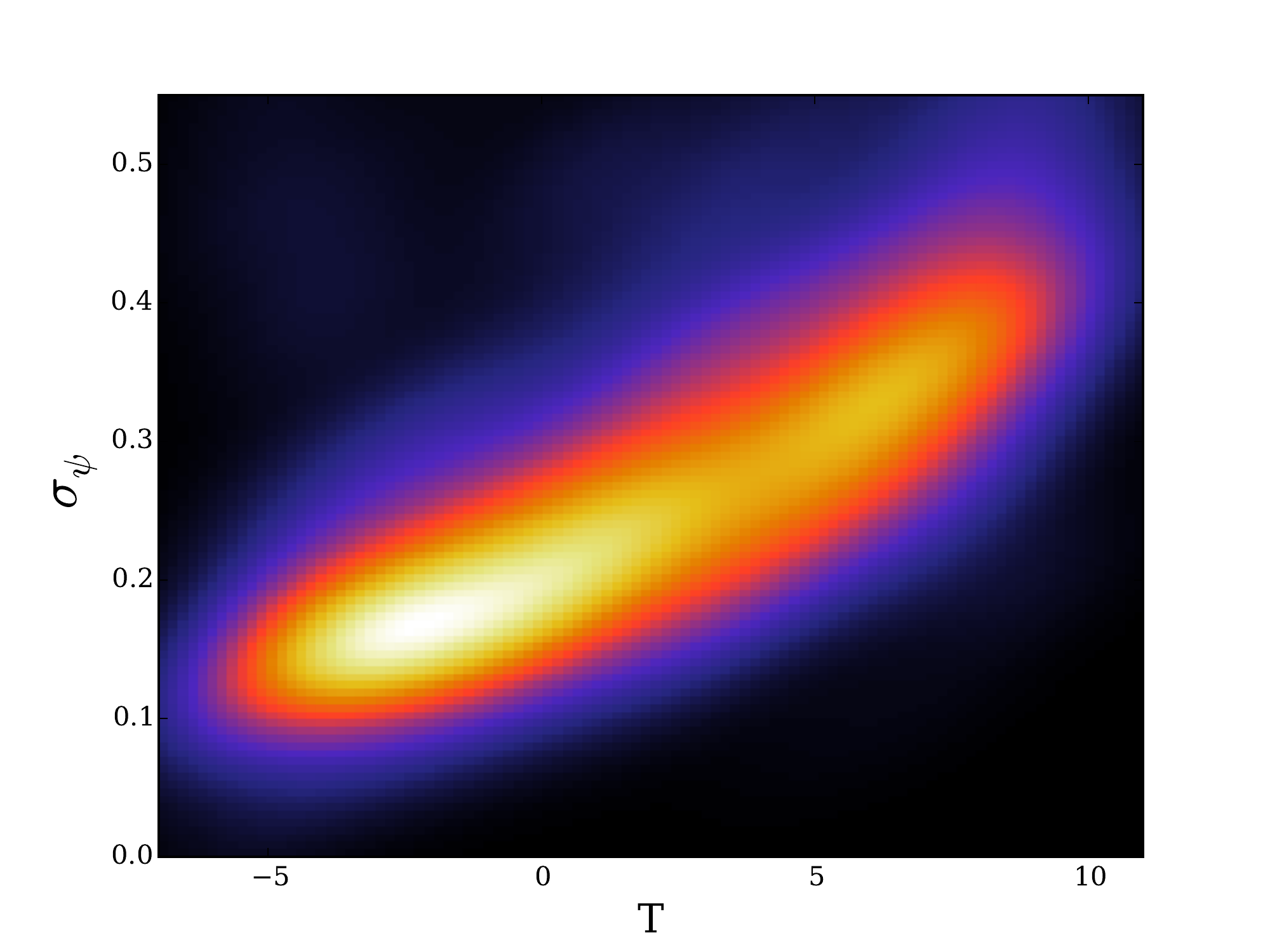}
	\caption{Density plot for $\sigma_\psi$ versus T-type for the
          EFIGI sample. For this plot, a subsample of galaxies were
          selected so that there is 45 galaxies in each T-type.}
\label{fig:density_T_sigma_psi}
\end{figure}

Figure \ref{fig:density_T_sigma_psi} shows a density plot of
$\sigma_\psi$ versus T-type for the EFIGI database (a subsample of
galaxies were selected such that there are 45 objects in each T-type),
where a clear linear relationship is seen. So, $\sigma_\psi$ is a good
diagnostic for T-type, provided there is enough spatial resolution to
distinguish spiral arms, rings and bars. This is the case for EFIGI
database, marginally for NA and not for LEGACY in general, as inferred
from the discussion in Section \ref{sec:feature_selection} and Figure
\ref{fig:feature_relative_importance}.

\section{The {\sc Morfometryka} algorithm}
\label{sec:morfometryka}

\begin{figure*}
\centering

\includegraphics[width=\linewidth]{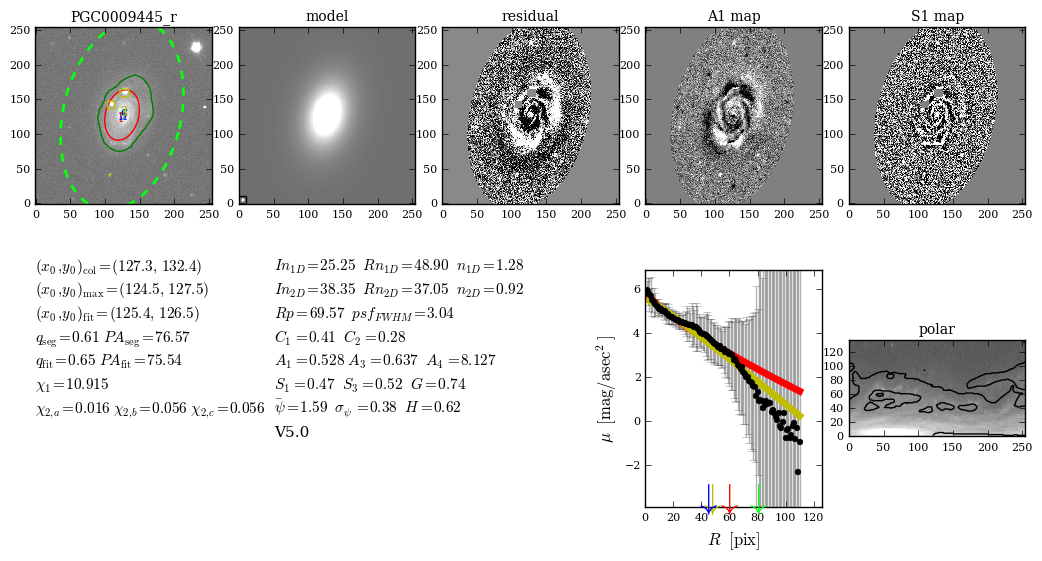}
\caption{Example \mfmtk\ graphical output for EFIGI $r$-band image of
  PGC 9445 (ObjID-DR7 587\,731\,513\,150\,734\,392). \textit{Top}, from left to
  right: \textbf{PGC0009445\_r}: original image (dark green line is
  primary segmentation, red line is 2D Sérsic fit $R_n$, dotted light
  green line is Petrosian region of 2 $R_p$, color caracters mark
  different centers); \textbf{model}: 2D Sérsic model image (insert
  showing the PSF at bottom left); \textbf{residual}: image minus 2D
  Sérsic model residual; \textbf{A1 map}: asymmetry map used to
  compute $A_1$; \textbf{S1 map}: Smoothness map used to
  $S_1$. \textit{Bottom:}, left to right: various measurements (see
  text for details); brightness profile (arbitrary units): black dots
  are measurements, red line 1D and yellow line 2D Sérsic
  fits. \textbf{polar}: map used to compute image gradient and
  $\sigma_\psi$. Its morphometric index is $M_{\rm i}=0.16$ }
\label{fig:example_output_mfmtk}
\end{figure*}

We developed a standalone application to automatically perform all the
structural and morphometric measurements over a galaxy image, called
\mfmtk\footnote{\texttt{http://morfometryka.ferrari.pro.br}}\ (\mfm).  \mfm\ reads the input stamp image and related PSF for
a given galaxy and performs various measurements explained in detail
in Appendix \ref{app:mfmtk}.  \mfm\ is currently implemented in an
object-oriented fashion in Python 2.7\footnote{Python Software
  Foundation. Python Language Reference, version 2.7. Available at
  http://www.python.org}, with the aid of scientific libraries SciPy
and Numpy (\cite{scipy}), Matplotlib (\cite{matplotlib}) and
PyFits\footnote{PyFits is a product of the Space Telescope Science
  Institute, which is operated by AURA for NASA}.

The \mfmtk \ basic output is: sky background value and standard
deviation; image centers $(x_0,y_0)_{\rm CoL}$, $(x_0,y_0)_{\rm
  peak}$; S\'ersic parameters for 1D surface brightness profile
fitting ($I_{n1D}; R_{n1D}; n_{1D} $) and for 2D image fitting
($I_{n2D}, R_{n2D}, n_{2D}, q_{2D}, PA_{2D}, (x_0,y_0)_{2D} $),
Petrosian Radius $R_p$; radii $R_{20}, R_{50}, R_{80}, R_{90}$ and
concentrations $C_1$ and $C_2$; asymmetries $A_1, A_2, A_3$ and fitted
center for $A_1$ and $A_3$; smoothness $S_1$ and $S_3$; Gini
coefficient $G$; second moment $M_{20}$; gradient field direction
value $\psi$ and standard deviation $\sigma_\psi$; quality flags QF.
Optionally, all maps (star masks, segmentation map, polar image and so
on) are saved. \mfmtk \ takes about 12 seconds to process a $256\times
256$ galaxy and $45\times 45$ PSF image on a 2.5 Ghz processor.  The
version used in this work was 5.0.

\section{Supervised  Classification}
\label{sec:classification}

Our morphological classification is based on the linear discriminant
method which separates galaxies in two main classes (E and S) in
morphometric parameters space. We train the classifier using the
classification from the Galaxy Zoo \citep{zoo1,zoo2}. The process was
done independently for EFIGI, NA, LEGACY and LEGACY$-zr$ datasets.

Our main goal is not only to classify galaxies in a way that
reproduces the human classification but also to establish a basis for
a morphometric space where galaxy classes are separated allowing
further studies where the human classifier cannot be used. Thus, we
use a linear discriminant and also we seek for the smallest set of
independent parameters that may yield a reliable classification which
is physically meaningful.

\subsection{Feature Selection}
\label{sec:feature_selection}

\begin{figure}
\centering
\includegraphics[width=\linewidth]{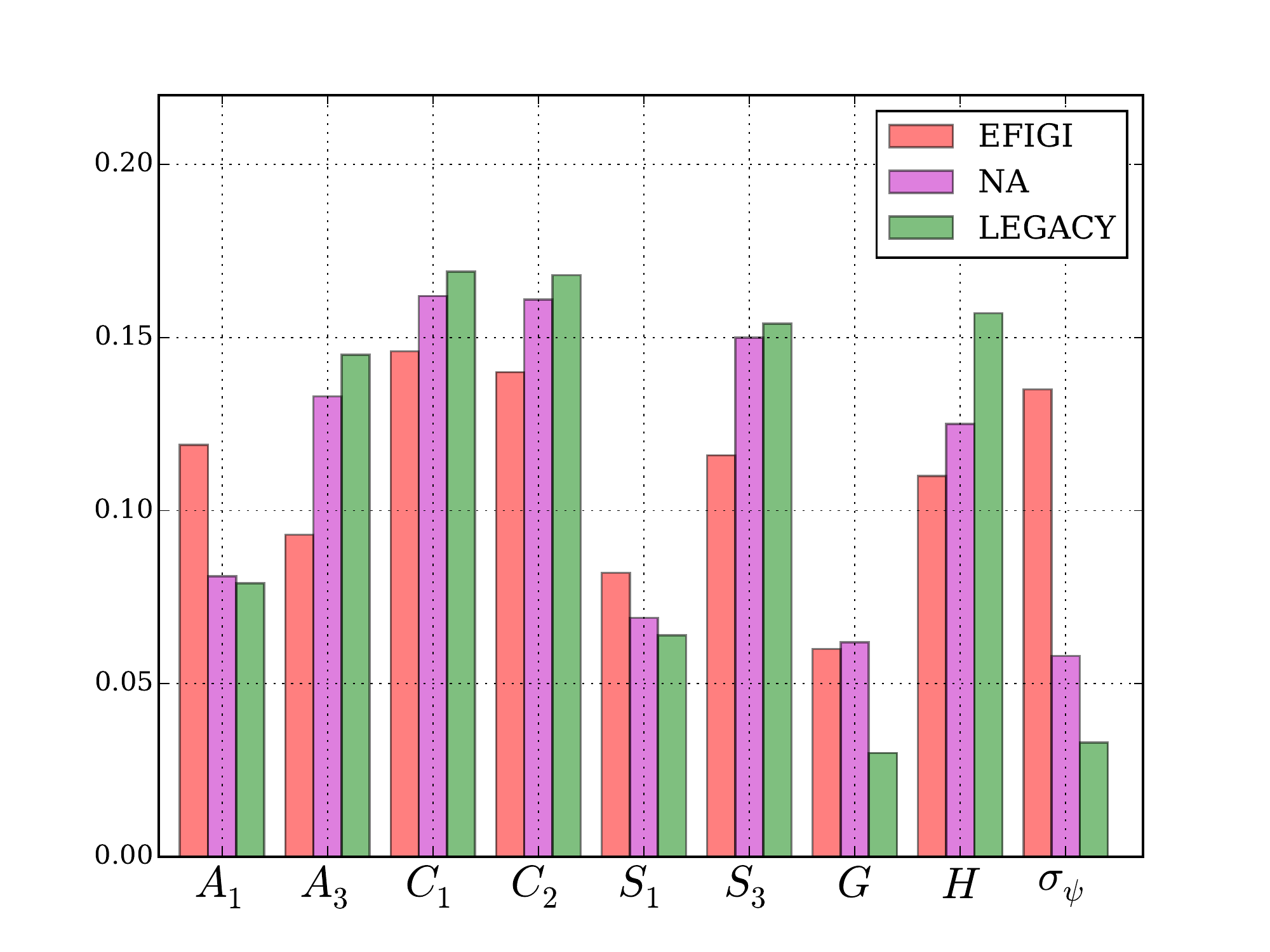}
\caption{Feature Relative Importance for the morphometric parameters
  as calculated by the Maximum Information Content criterion.}
\label{fig:feature_relative_importance}
\end{figure}

Given that we have so many measured quantities for each galaxy, some
of them may be redundant or irrelevant and we need to select those which are
more relevant to the classification algorithm. Many feature selection
algorithms tends to diminish the importance of quantities that
correlate with each other. A criterium that avoids that is the Maximum
Information Content (MIC, \cite{MINE}). MIC is based on the mutual
information and the information entropy: it compares, given the
parameters and the known class, which one possesses the greater mutual
information with the class variable, i.e. which one will have greater
impact in the classification.  The normalized values for MIC are shown
in Figure \ref{fig:feature_relative_importance}.
 
We may have more information on how efficiently each feature helps to
separate the classes by examining the feature histogram separated by
classes, as shown Appendix \ref{app:histogramas}, in Figure
\ref{fig:EFIGI_ima_r_v47_T_distrofit} (EFIGI),
Fig. \ref{fig:Abraham_stamps_v47_NAtable2_distrofit} (NA),
Fig. \ref{fig:LEGACY_run01_zoo_distrofit} (LEGACY) and
Fig. \ref{fig:LEGACY_run01_zoo_z01r1778_distrofit} (LEGACY$-zr$) .  We
must be aware that we are seeing marginal probability distribution
functions (PDF) on each variable and this is not equivalent to analyse
the multivariate PDF of all parameters together.

First, we note that the features with highest discriminant power are
those related to the light concentration (S\'ersic $n$, $C_1$ and
$C_2$). Since they are equivalent (Appendix \ref{app:C_and_n}), and
also equivalent to the Petrosian Radii (Appendix \ref{app:Rp_and_n}),
we retain only $C_1$ and $C_2$ which are not parametric and more
robust.

Comparing the asymmetry measures in the histograms in Appendix
\ref{app:histogramas}, we see that $A_3$ is able to discriminate
classes better than $A_1$, which is confirmed by the MIC values. Gini
coefficient is very poor at separating E from S, as it is
$M_{20}$. Compared to Gini, entropy $H$ works better in separating
classes, and the introduced $S_3$ is better than the original
$S_1$. The axis ratio $q$ is good for E but indifferent for S, so it
is not used. The spirality $\sigma_\psi$ is also good to discriminate
classes but it is crucially dependent on angular resolution - its
importance decreases from EFIGI to NA, to LEGACY, in the same sense as
the mean angular resolution decreases.

Finally, based on the MIC analysis, we choose this set of parameters
\[
\textbf{x} =\left\{C_1, A_3, S_3, H, \sigma_\psi   \right\},
\]
for they constitute a minimal set of independent parameters that yield
a reliable classification.  Four of the chosen parameters are new,
used here for the first time. $A_3$ and $S_3$ are enhanced versions of
standard parameters, $H$ is first applied in morphometric studies and
$\sigma_\psi$ is completely new.

\subsection{Linear Discriminant Analysis}
\label{sec:LDA}

A simple linear classifier may be represented by a discriminant
function which for a given input vector \textbf{x} that contains $d$
morphometric measurements \citep{Duda,Bishop} gives
\begin{equation}
\label{eq:discrfunc}
f(\textbf{x}) = \textbf{w}^T \textbf{x} + w_0,
\end{equation} 
\noindent where $\textbf{w}$ is the weight vector and $w_0$ is the
threshold. The input vector is assigned to the class $\mathcal{C}_1$
if $f(\textbf{x})>0$ and to $\mathcal{C}_2$ otherwise. The decision
boundary or surface is a hyperplane defined by $f(\textbf{x})=0$, for
which $\textbf{w}$ is a normal vector and $-w_0/||\textbf{w}||$ its
normal distance to the origin.  The decision function corresponds to
the perpendicular distance from $\textbf{x}$ to the decision surface.

When using the Bayes Decision Theory the expressions for $\textbf{w}$
and $w_0$ are assigned as follows: an object belongs to class
$\mathcal{C}_1$ if
\begin{equation}
\label{eq:bayes1}
P(\mathcal{C}_1,\textbf{x}) > P(\mathcal{C}_2,\textbf{x}) \qquad (\text{for class }\mathcal{C}_1)
\end{equation}
and to $\mathcal{C}_2$ otherwise. Since the evidence $p(\textbf{x})$
is the same for both classes, Bayes rule in Eq. (\ref{eq:bayes1}) is
equivalent to
\begin{equation}
\label{eq:bayes2}
p(\textbf{x}, \mathcal{C}_1) \, P(\mathcal{C}_1 )  >  p(\textbf{x}, \mathcal{C}_2) \, P(\mathcal{C}_2 ) 
\end{equation}
\noindent where $p(\textbf{x}, \mathcal{C}_i)$ is the class
conditional probability density function (CCPDF) and
$P(\mathcal{C}_i)$ the prior. We assume that the CCPDF is multivariate
normal density
\begin{equation}
  p(\textbf{x}, \mathcal{C}_i)= \frac{1}{(2\pi)^{d/2}
    |\pmb{\Sigma}|^{1/2}} \exp\left[ -\frac{1}{2} (\textbf{x} -
    \pmb{\mu}_i)^T \, \pmb{\Sigma}_i^{-1} (\textbf{x} - \pmb{\mu}_i)
    \right],
\end{equation}
\noindent where $\pmb{\mu}_i$ are the mean and $\pmb{\Sigma}_i$ the
covariance matrix of $\textbf{x}$ for class $\mathcal{C}_i$ .  The
decision rule Eq. (\ref{eq:bayes2}), or equivalently its logarithm, is
then
\begin{eqnarray}
\label{eq:bayes3}
-\frac{1}{2} (\textbf{x} - \pmb{\mu}_1)^T \, \pmb{\Sigma}_1^{-1}
(\textbf{x} - \pmb{\mu}_1) + \ln P(\mathcal{C}_1) > \nonumber
\\ -\frac{1}{2} (\textbf{x} - \pmb{\mu}_2)^T \, \pmb{\Sigma}_2^{-1}
(\textbf{x} - \pmb{\mu}_2) + \ln P(\mathcal{C}_2)
\end{eqnarray}
The terms involving $\textbf{x}'\boldsymbol{\Sigma}_i^{-1}\textbf{x}'$
are general quadratic forms and if we expand them in
Eq.(\ref{eq:bayes3}) we have a Quadratic classifier. But instead, we
consider identical covariance matrices
$\boldsymbol{\Sigma}_1=\boldsymbol{\Sigma}_2=\boldsymbol{\Sigma}$
which yield a Linear classifier. Expanding the terms in
Eq.(\ref{eq:bayes3}), and ignoring those that are identical for both
classes, we have
\begin{align}
\label{eq:bayes4}
\boldsymbol{\Sigma}^{-1}(\boldsymbol{\mu}_1 - \boldsymbol{\mu}_2) +
\frac{1}{2}\boldsymbol{\mu}_1^T \boldsymbol{\Sigma}^{-1}
\boldsymbol{\mu}_1 + \frac{1}{2}\boldsymbol{\mu}_2^T
\boldsymbol{\Sigma}^{-1} \boldsymbol{\mu}_2 + \ln
\frac{P(\mathcal{C}_1)}{P(\mathcal{C}_2)}>0.
\end{align}
If we refer to Eqs.\ref{eq:discrfunc} and \ref{eq:bayes4}, we have
then
\begin{align}
\label{eq:lda_w}
\textbf{w} &= \boldsymbol{\Sigma}^{-1}(\boldsymbol{\mu}_1 -
\boldsymbol{\mu}_2)\\ 
\label{eq:lda_w0}
w_0 &= \frac{1}{2}\boldsymbol{\mu}_1^T
\boldsymbol{\Sigma}^{-1} \boldsymbol{\mu}_1 +
\frac{1}{2}\boldsymbol{\mu}_2^T \boldsymbol{\Sigma}^{-1}
\boldsymbol{\mu}_2 + \ln \frac{P(\mathcal{C}_1)}{P(\mathcal{C}_2)},
\end{align}
\noindent which completes our linear classifier.

\subsection{Classifier Performance}
\label{sec:performance}

We may \rrbf{estimate} the classifier performance by means of a
\textbf{confusion matrix} or contingency table, which is a comparison
of the actual class with the predicted class for each objects. The
performance is \rrbf{evaluated} calculating several \textbf{scores} based on
true positives (TP, hits), true negatives (TN, correct rejections),
false positives (FP, false alarms) and false negatives (FN,
misses). See for example \cite{hackeling}.

The accuracy $A$ is the fraction of hits relative to the total number of classifications 
\[
A = \frac{TP+TN}{TP+TN+FP+FN};
\]
Precision $P$ is the fraction of positive predictions that are correct 
\[
P = \frac{TP}{TP+FP};
\] 
Sensitivity $R$ is the fraction of the truly positive instances that the classifier recognizes
\[
R=\frac{TP}{TP+FN};
\]
and $F_1$ score which is harmonic mean between sensitivity and precision
\[
F_1 = \frac{2TP}{2TP + FP + FN}.
\]

We test the performance of the classifier by a 10-fold cross
validation: for each database we selected those samples with known
Galaxy Zoo classification and partitioned them in 10 parts; in each of
10 runs one of the parts we used as a validation sample and the other
9 parts as training sample. In each run, the scores A, P, R, F1 were
calculated; their final averages are shown in Table \ref{tab:scores}.
For all databases the classifier performs usually better than 90\%,
\rrbf{namely 90\% of the time the automated classifier grees with the
  visual classification.  If we consider that the performace in the
  human classification is also of that order,} and that those
classification were used to train the classifier, then this
performance can be considered very good and it is the best figure that
we could expect without using a classifier that would incorporate the
errors in it.

\begin{table}
\begin{tabular}{|l|r|r|r|r|}
\hline               & EFIGI    & NA     & LEGACY       & LEGACY--$zr$ \\ \hline  
$A$                  & 0.938    & 0.902   & 0.877      & 0.938        \\
$P$                  & 0.962    & 0.931   & 0.905      & 0.956       \\
$R$                  & 0.964    & 0.899   & 0.935      & 0.968        \\
$F_1$                & 0.963    & 0.914   & 0.920      & 0.963        \\ \hline
\end{tabular} 
\caption{Mean scores for each database for a 10-fold cross validation tests.}
\label{tab:scores}
\end{table}

\section{Morphometric Index}
\label{sec:Mi}
	
\begin{figure*}
	\centering
	\includegraphics[width=\linewidth]{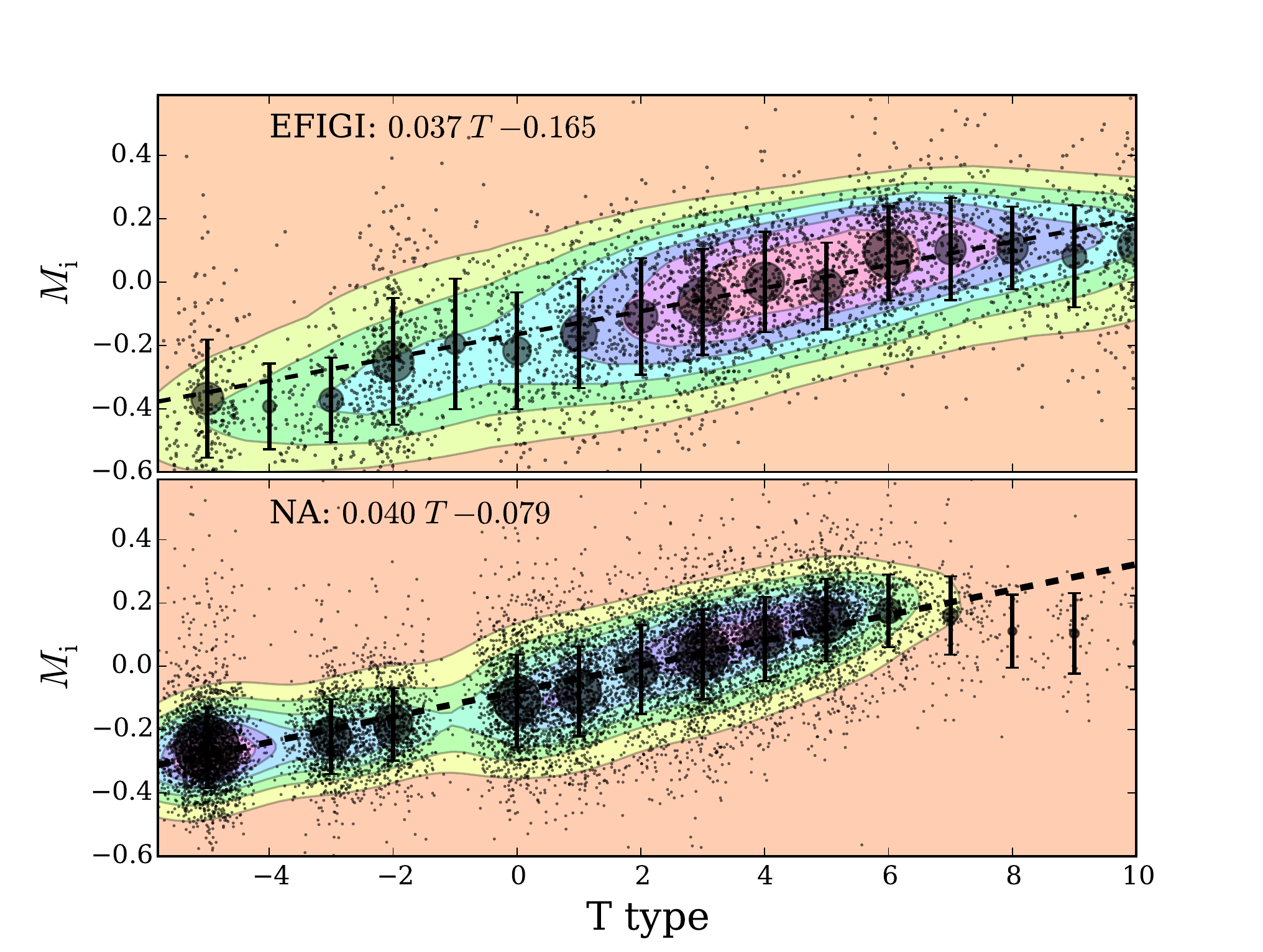}
	\caption{The relationship between the morphological T-type and
          the morphometrix index $M_{\rm i}$ for EFIGI (above) and NA
          (below) databases. Small solid dots are individual galaxies;
          large circles indicate mean $M_{\rm i}$ for each integer
          T-type and their size is proportional to number of object;
          error bars indicate standard deviation in $M_{\rm
            i}$. Contours are draw for a kernel representation of
          points.  The dashed line are the best linear regression,
          whose parameters are shown on top of each plot. }
	\label{fig:morphometric_index}
\end{figure*}

As stated in Section \ref{sec:LDA}, the discriminant function
$f(\textbf{x})$ is the distance of $\textbf{x}$ to the plane that
separates classes, here ellipticals and spirals. Based on that, we
propose to use $f(\textbf{x})$ to represent the galaxy type, that we
call the \textbf{morphometric index} $M_{\rm{i}}$.  Figure
\ref{fig:morphometric_index} shows the comparison of $M_{\rm{i}}$ with
the $T$ type from EFIGI and NA samples. There is a clear linear
relationship between $M_{\rm{i}}$ and $T$ type justifying the use of
$M_{\rm{i}}$ as a morphometric index. In Section \ref{sec:results} we
extend this argument by comparing $M_{\rm i}$ with other galaxy
physical characteristics. By construction, $M_{\rm i}$ is negative for
early-type and positive for late-type galaxies.

A linear regression between $T$ and $M_{\rm i}$ could be used to
calibrate $M_{\rm i}$ as an inferred $T$-type, but since $T$ is a
subjective parameter we prefer to maintain $M_{\rm i}$ in its own
scale, and as a pure morphometric measure, estimated solely based on
the values of $\textbf{x}$. For a binary classification, the magnitude
of the direction vector $\textbf{w}$ has no importance, and in fact,
$\|\textbf{w}\|$ could be different depending on the details of the
LDA. But since we want to use this distance as a physical measure, we
prefer to normalize $\textbf{w}$ and the distance to the plane $w_0$
in the morphometric index $M_{\rm i}$, so that
\begin{align}
M_{\rm i} &= \hat{\textbf{w}}^T \textbf{x} + \hat{w_0},
\\ \text{where}\quad \hat{\textbf{w}} &=
\frac{\textbf{w}}{\|\textbf{w}\|} \quad \text{and}\quad \hat{w_0} =
\frac{w_0}{\|\textbf{w}\|}. \nonumber
\end{align}  
The final values for LEGACY database are
\begin{align}
\hat{\textbf{w}} &= \{-0.832, 0.249, 0.451, 0.190,  -0.079\} \\
\hat{w}_0  &= 0.018
\end{align}
	
As we see in Eq.(\ref{eq:lda_w0}), $w_0$ depends on the class priors
$P(\mathcal{C}_i)$. Usually, the relative frequency for each class
$N_i/N$ ($N_i$ the number of objects in class $\mathcal{C}_i$, $N$
total number of objects) is used as the prior, since it gives the
probability that a new object belongs to class $\mathcal{C}_i$ if we
know nothing about it. But in the EFIGI and NA databases, the relative
frequency is biased, as it was designed to contemplate each
morphological T type with approximately the same number of
objects. So, the priors and hence $w_0$ for EFIGI and NA could not be
applied to other databases with different relative frequencies. LEGACY
and LEGACY--$zr$ databases, on the contrary, have priors that may
reflect real distribution of classes of galaxies, since no morphology
was used to select the objects.  Briefly, the intercept term in the
linear relationships in Figure (\ref{fig:morphometric_index}) may be
biased by the selection effects in the databases; the linear character
however is unaffected.
		
The linear regression between T-type and $M_{\rm i}$, shown in
Fig.(\ref{fig:morphometric_index}) is a least square solution using a
robust Theil--Sen estimator, which computes the median slope among all
pairs of points in a set, impemented in Scikit-Learn library
\citep{scikit-learn}. Note that $T \simeq 20\, M_{\rm i}$.
	
\begin{figure}
\centering
\includegraphics[width=0.5\textwidth]{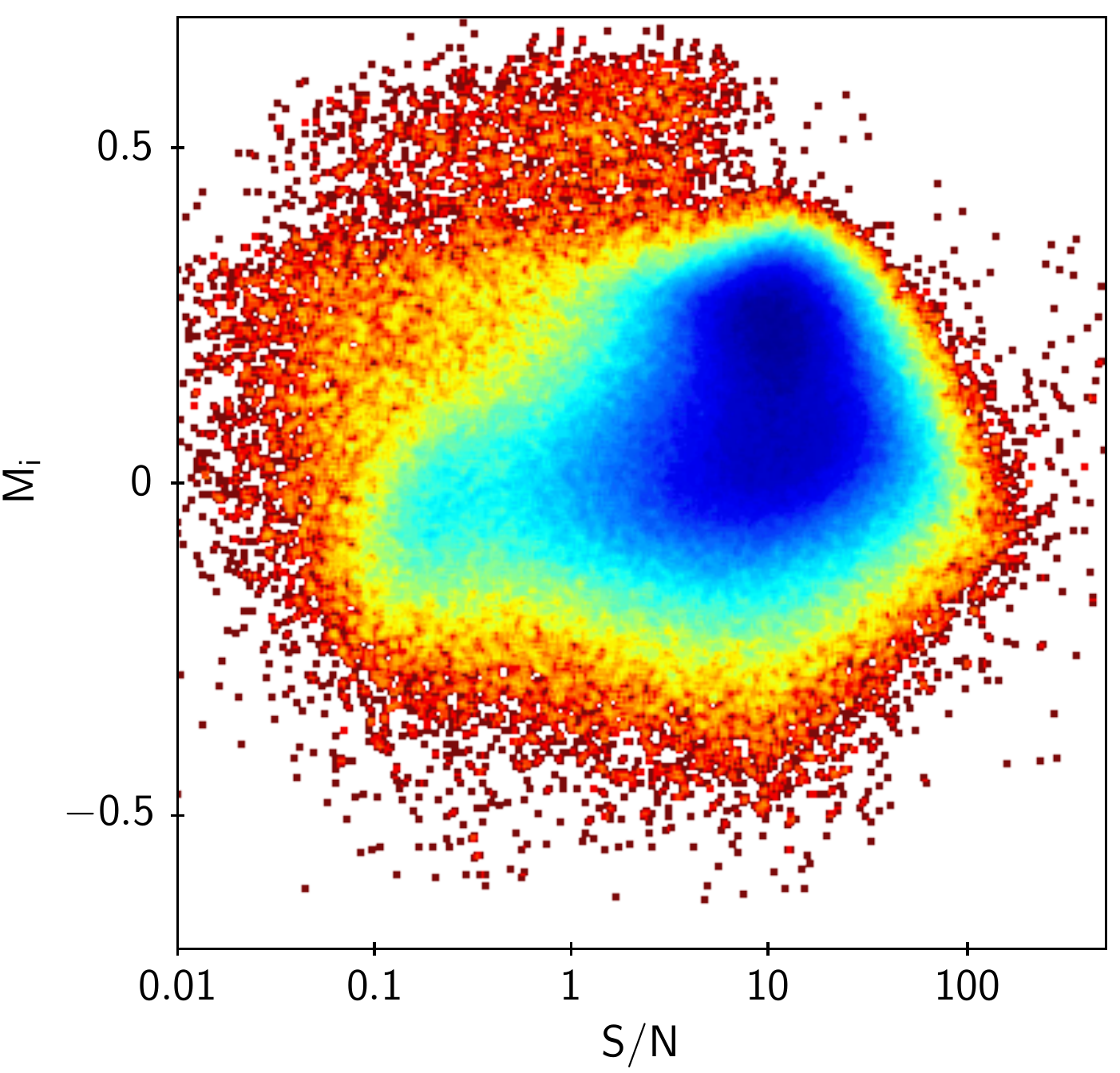}
\caption{Comparison between the signal-to-noise of the galaxy defined
  as S/N$=I_{n,2D}$/\texttt{skybgstd} and the Morphometric Index
  M$_{\rm i}$.}
\label{fig:Fig_Mi_SN}
\end{figure}

In order to have M$_{\rm i}$ as a trustable morphological indicator we
need to establish how accurate it is, which in principle can be done
by propagating the error from each of the parameters $C_1$, $A_3$,
$S_3$, $H$ and $\sigma_\psi$. However, we find it more realistic to
compute the signal-to-noise of the galaxy as
S/N$=I_{n,2D}$/\texttt{skybgstd} (see Appendix \ref{app:mfmtk}) and
see how M$_{\rm i}$ varies with it. As we can see from Figure
\ref{fig:Fig_Mi_SN}, there seems to be no trend between them and the
blue area indicates that most galaxies have S/N around 10 and average
M$_{\rm i}$ slightly around 0.1, which is slightly above 0.0, where we
would expect.
	
\section{Comparison with other physical parameters}
\label{sec:results}

\begin{figure*}
	\centering
	\includegraphics[width=0.9\textwidth]{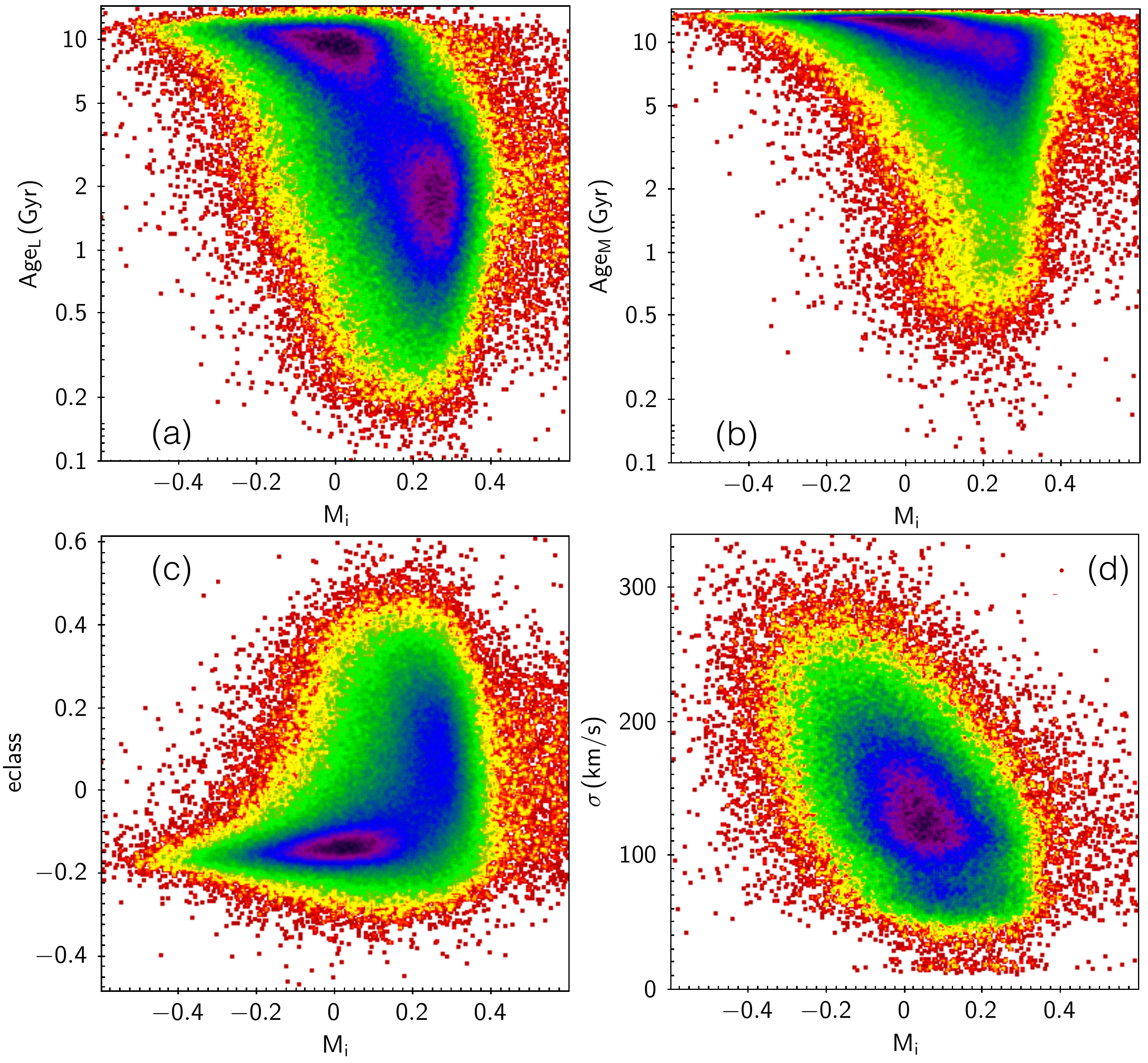}
	\caption{Relationship between M$_{\rm i}$ and Age$_{\rm L}$
          (age weighted by luminosity), Age$_{\rm M}$ (age weighted by
          mass), eClass and central velocity dispersion $\sigma$}
	\label{fig:populacao_estelar_relacoes}
\end{figure*}

We present  in this paper a new approach for galaxy morphological
classification that is not focused on recovering visual
classification, although this is done remarkably well (see Section
\ref{sec:performance}).  In this work, the parameters defining the morphology
of a galaxy are physically motivated and to confirm how successful we
were in reaching this goal we compare $M_{\rm{i}}$ with quantities
measured from the spectrum of the galaxies we measured.

Figure~\ref{fig:populacao_estelar_relacoes} exhibits how Age$_{\rm L}$
(age weighted by luminosity), Age$_{\rm M}$ (age weighted by mass),
eClass (a single parameter classifier based on PCA analysis, retrieved
from the SDSS database), and central velocity dispersion $\sigma$
correlate with $M_{\rm{i}}$. Notice that the SDSS spectra reflect
properties of the central region of galaxies. In
Figure~\ref{fig:populacao_estelar_relacoes}.a we see that, overall,
negative $M_{\rm{i}}$ corresponds to older systems. Age$_{\rm L}$
reflects more recent episodes of star formation and in this case as
$M_{\rm{i}}$ goes to very negative the systems do not present any
recent star formation, namely these are very old galaxies. There is a
ridge of old systems extending from $M_{\rm{i}}$ = -0.5 up to 0.1 and
then a significant drop in age as $M_{\rm{i}}$ tends to 0.2. For $0.1
< M_{\rm{i}} < 0.4$ we see that Age$_{\rm L}$ is around 1.5 Gyr. These
are the late type spirals which exhibit a considerable amount of star
formation. This figure clearly shows that morphology, in this case
manifested by the parameter $M_{\rm{i}}$, varies continuously from old
to young stellar population which is an important aspect of any
morphological quantifier -- to reflect stellar population properties
of galaxies. 

Figure~\ref{fig:populacao_estelar_relacoes}.b is similar
to Figure~\ref{fig:populacao_estelar_relacoes}.a but plotting
Age$_{\rm M}$ instead, which contrary to Age$_{\rm L}$, reflects more
the whole star formation history of a galaxy. The same trend is seen
here, however since the SDSS spectra samples only the central region
of the galaxies, there is a population that dominates the figure for
Age$_{\rm M}$ around 10.0 Gyr, from early (negative $M_{\rm{i}}$) to
late types (positive $M_{\rm{i}}$). 

Figure~\ref{fig:populacao_estelar_relacoes}.c exhibits how
$M_{\rm{i}}$ is related to eClass, a parameter designed to express
differences in stellar populations among different galaxies and then
serving as a discriminant between early and late type systems. We find
that the relation between these two quantities is not linear, which is
what we would expect if both reflected morphology in a one-to-one
relationship. What we see is that for $-0.5 < M_{\rm{i}} < 0.2$ eClass
is concentrated around -0.15 (early type systems), with a scatter that
increases as $M_{\rm{i}}$ increases. Then for $M_{\rm{i}} > 0.2$
eClass increases steadily reaching eclass of 0.5 for $0.2 < M_{\rm{i}}
< 0.4$.  \rrbf{ Both eClass and $M_{\rm i}$ are associated to
  morphology, although eClass is primarily associated with stellar
  population and $M_{\rm i}$ is derived solely based on image
  morphometry. $M_{\rm i}$ is more sensitive to morphology,
  particularly in the early-type systems domain ($M_{\rm i}<0$).}

Finally, in Figure~\ref{fig:populacao_estelar_relacoes}.d we present
the relation with the central velocity dispersion $\sigma$ (corrected
for an aperture of $R_{e}/8$, where $R_{e}$ is the effective radius of
the galaxy). Even though with a large scatter a clear relation exists
between $M_{\rm{i}}$ and $\sigma$, which is remarkable considering
that $M_{\rm{i}}$ is solely photometric. In summary, these comparisons
show that $M_{\rm{i}}$ is reliable in separating the different
morphological types according to their stellar population properties,
a performance not seen in other previously proposed morphological
quantifiers.

\section{Summary}
\label{sec:summary}

We present a new method to establish morphological classification of
galaxies that is physically motivated although it matches what is done
visually for the very nearby Universe equally well. In the following,
we summarize the main aspects of the classification system proposed
here and the verification analysis:

1 - We developed a pipeline that automatically estimates morphometric
parameters from galaxy images. Measured parameters include
Concentration, $C_1$, Asymmetry, $A_3$, and Smoothness, $S_3$ which
were slightly modified with respect to the conventional ones. We also
make use of two new extra parameters: entropy $H$ and spirality
$\sigma_\psi$.
 
2 - \mfmtk\ measures several quantities per galaxy which brings the
question of which ones are more adequate for establishing the
morphological type of the system. We use a method called Maximum
Information Content (MIC) to select the relevant features avoiding
redundancy. The new introduced morphometric parameters have a better
discriminant power than previously used ones.  MIC analysis resulted
in the minimum number of independent parameters listed in item 1. The
relationship between concentration, Petrosian radius and Sérsic index
$n$ is derived in Appendix \ref{app:C_and_n} and \ref{app:Rp_and_n}.

3 - Our supervised classification is based on Galaxy Zoo and tested
with different datasets: EFIGI, NA, LEGACY and LEGACY--$zr$. The
Linear Discriminant Analysis (LDA) method is used to determine the
decision surface that separates early from late type systems and the
distance from this surface will indicate how early or late the system
is. It is exactly this distance that we propose as a morphological
index, $M_{\rm{i}}$.

4 - Classification performance was evaluated using the confusion
matrix, from which we measured accuracy, precision and sensitivity
scores, with a 10-fold cross validation scheme. We obtain final
scores better than  90\%.

5 - Another independent validation comes from comparing $M_{\rm{i}}$
with stellar population quantities and velocity dispersion which were
established using the spectra available in DR7 together with the
spectral fitting code \textsc{starlight}. We note that $M_{\rm{i}}$ correlates
with eClass and it shows that classifying early-type galaxies solely
as eClass $< 0$ can significantly contaminate the sample with
late-type systems which have $M_{\rm{i}} > 0.2$.

\acknowledgments 

\rrbf{We thank the referee Ronald Buta for many suggestions that
  helped improve the manuscript.}  We also thank Karina Machado and
Diana Adamatti (C3--FURG) for lending us their processing
infrastructure; Juliano Marangoni (IMEF--FURG) for 
discussions on statistics; Val\'erie de Lapparent (IAP--Paris) for help with EFIGI data.
RRdC would like to thank Francesco La Barbera for insightful
discussions on the topic over the years.  RRdC acknowledges financial
support from FAPESP through a grant \# 2014/11156-4.

\appendix

\section{Morfometryka Algorithm Details}
\label{app:mfmtk}

Here, we provide a detailed description of the various measurements in
\mfmtk.  \mfm\ is logically divided into four main blocks (classes in
programming parlance): {\sc Stamp} -- basic data reading and low
level, low complexity geometrical measurements; {\sc Photometry}~--~
luminosity distribution, star masking and Petrosian radius estimation;
{\sc S\'ersic}~--~1D and 2D luminosity distribution fitting; {\sc
  Morphometry}~--~measurements of the morphometric parameters used
later on for establishing the galaxy's morphology.  The package also
include auxiliary applications \textsc{makemySDSS} for retrieving SDSS
frames and cutting stamps and \textsc{LDAclassify} to perform the
Linear Discriminant Analysis.  In the following, logical units are
written in {\sc small caps}, algorithm code in {\tt typewriter}.
	
\subsection{Cutting stamps}

The list of all objects, containing \texttt{ObjIDs, RA, DEC, run,
  rerun, camcol, field} and \texttt{petroRad}, is generated with the
following SQL query on SDSS CasJobs

\begin{verbatim}	
SELECT	p.objID, p.ra, p.dec, p.run, p.rerun, p.camcol, p.field, p.petroRad_r 
	FROM DR7.SpecObj as s JOIN DR7.PhotoObj AS p ON s.bestObjID = p.objID  
	WHERE s.specclass = 2.	
\end{verbatim}
From this list, we build a set of unique combinations of
\texttt{(run,rerun,camcol,field)} and the required SDSS Frames and
psFields are downloaded. We do it exactly as for DR7 Frames and
psFields but we download the DR10 files since they refer to the same
region of the sky, i.e. the raw data are the same, but the image
processing algorithms were improved from DR7 to DR10. Also, DR10
frames are calibrated in
nanomaggies\footnote{https://www.sdss3.org/dr10/algorithms/magnitudes.php}
\citep{Lupton}. For each object, the relative Frame is loaded and a
square region of size 10 \texttt{petroRad\_r} centered in the object's
RA and DEC is cut. The PSF for the same position is generated with the
SDSS \texttt{read\_PSF} application from the \texttt{psField}
file. The stamp FITS file header is updated with the astrometry and
relevant frame keywords. If the object is in the frame border, i.e.,
if it has less than 90\% of pixels in the frame, a FITS header
keyword \texttt{FLAGINC} and a header comment "\texttt{MFMTK:
  incomplete stamp}" is written.

\subsection{Basic image processing} 
\label{sec:stamp}

The process starts with the target image \texttt{gal0} and the
associated PSF, which is measured from the second moment collapsed in
the $y$ direction.The \textbf{sky background} \texttt{skybg} is
estimated from the median of all pixels from the four corners of the
image (squares of typical width of 10 pixels,
\texttt{skyboxsz}).\footnote{The median estimate is less affected than
  the mean by outliers, and such sky background estimate has proven to
  be accurate enough even if a star occupies several pixels in one of
  the corners.} The accuracy of the sky background estimate
\texttt{skybgstd} is set by the standard deviation of the
aforementioned set of pixels.

The \textbf{segmentation} is done on the \texttt{gal0fltr} image,
which is the \texttt{gal0} image median filtered with a window of size
\texttt{segS} (typically 5 pixels); high frequencies are filtered from
the image to avoid sharp edges in the segmented regions. Regions are
then selected by \textbf{histogram thresholding}: those pixels whose
intensity are greater than the threshold {\sf
  median}(\texttt{gal0fltr})+\texttt{segK}$\cdot$ {\sf
  mad}(\texttt{gal0fltr}) are selected, where \textsf{mad} is the
median absolute deviation.  This threshold selects \texttt{segK}
\textsf{mad} above the median, which is similar to sigma-clipping $K$
standard deviations above the mean, except that median and the median
absolute deviation (\textsf{mad}) are used, which are more robust to
outliers and intensity variations. This histogram thresholding
operates on the intensity space only and may select regions that are
not spatially connected. The spatial information is taken into account
by performing a \textbf{connected-component labeling}, where
4--connected pixels receive the same label. At this stage the
segmentation consists of one or more labelled regions. The final
segmented region is then selected either by size (largest) or by
position (center of light closest to image center), depending on
configuration. For SDSS stamps the position criteria is used.  A
segmentation mask \texttt{segmask} is made from the selected region,
from which a segmented galaxy image \texttt{gal0seg} is derived; on
both, pixels outside the segmentation region are nil. Geometric
measurements in this section are done in the \texttt{gal0seg} image.

The galaxy image center is estimated in two distinct ways. First, the
\textbf{peak center} $(x_0,y_0)_{\rm peak}$, referring to the locus
where the intensity is maximum, is estimated from the center of light
(first moment) of the 5$\times$5 matrix around the pixel with highest
intensity, attainning sub-pixel precision.

For an image $I(x,y)$  the standard image moments are defined as 
\[
m_{pq}=\int\limits_{-\infty}^{\infty} \int\limits_{-\infty}^{\infty} x^py^q I(x,y) \,dx\, dy,  
\]
and the \textbf{center of light} (CoL) $(x_0,y_0)_{CoL}$ are given by
$x_0 = m_{10}/m_{00}$ and $y_0 = m_{01}/m_{00}$. The translational
invariant moments $\mu_{pq}$ are determined by replacing $x$ and $y$
by $(x-x_0)$ and $(y-y0)$, respectively, in $m_{pq}$.  The axis
lengths are given by
\begin{align}
\lambda_1 = \frac{\sqrt{|\mu_{20}+\mu_{02} + \Lambda |}}{m_{00}} \qquad
\lambda_2 = \frac{\sqrt{|\mu_{20}+\mu_{02} - \Lambda |}}{m_{00}}  \\
\text{where}  \qquad
\Lambda = \sqrt{(\mu_{20}-\mu_{02})^2 + 4 \mu_{11}^2}
\end{align}
from which we define
\begin{align*}
a = \max{(\lambda_1,\lambda_2)} \\
b = \min{(\lambda_1,\lambda_2)}.
\end{align*}
Further, we can calculate the position angle of the main axis by
\[
PA = \frac{1}{2}\  \arctan(2 \mu_{11}, \mu_{20}-\mu_{02}).
\]
Details of the derivation of the relations above are given, for
example, in \citep{Flusser}.

For future use, a standardized version of the segmented galaxy image
is calculated: given the parameters estimated above, an affine
transform is applied to \texttt{gal0seg} such that in the resulting
\texttt{stangal} image the object is centered in the image array, has
zero position angle and its axial ratio is unity. Optionally the
integrated luminosity can be normalized.

\subsection{Photometry Routines}
	\label{sec:photom}
	
Photometry is performed by positioning successive ellipses relative to
a fixed center $(x_0,y_0)_{\rm peak}$, with constant axis ratio $b/a$
and position angle $PA$. The photometric measurements are performed
for ellipses with semi-major axis $R$ ranging from 1 pixel up to the
size of the image diagonal, in steps of 1 pixel. Later the profiles
are truncated (see below).
	
The pixels 1 pixel away from the ellipse are called border pixels
(\texttt{ellindxbrdr}) and the pixels inside the ellipse are internal
pixels (\texttt{ellindxin}).  In a given semi-major axis $R$, for the
set of border pixels, those pixels whose intensity are above some
threshold relative to the pixel group (given by {\sf
  median}(\texttt{ellindxbrdr})+\texttt{StarSigma}$\cdot${\sf
  mad}(\texttt{ellindxbrdr})) are masked as stars. The ellipse mean
intensity $I(R)$ and associated error $I_\mathrm{err}(R)$ are
calculated by the average and the standard deviation of the border
pixels not masked as stars. The total luminosity $L(R)$ is the sum of
the internal pixels for each ellipse, and the mean intensity $\langle
I \rangle (R)$ is given by the ratio of $L(R)$ and the number of
pixels inside the ellipse.
	
At each semi-major axis iteration, the Petrosian function,
Eq.(\ref{eq:petrosian_eta}), is evaluated and once it falls below the
critical value $\eta_0=5$, the Petrosian radius $R_p$ is evaluated by
linear interpolating $\eta(R)$ between the adjacent points.  A
Petrosian Region is defined as an elliptic region of semi-major axis
$N_{\!R_p} \cdot R_p$ (we use $N_{\!R_p}=2$) and the same axis-ratio
and position angle measure in the segmented image (Section
\ref{sec:stamp}), and stored as \texttt{petromask}.  The image
\texttt{galpetro} = \texttt{petromask} $\cdot$ \texttt{gal0} is
defined. The profiles $I(R)$, $I_{\rm err}$, $L(R)$ and $\langle I
\rangle (R)$ are cut at $N_{\!R_p} \cdot R_p$.

\subsection{The S\'ersic routines}

Standard 1D S\'ersic parameters are measured by fitting the S\'ersic law \citep{Sersic} 
\begin{equation}
\label{eq:leisersic}
I(R) = I_n \exp\left[-b_n\left(\frac{R}{R_n}\right)^{^1/_n} -1\right]
\quad \text{with} \quad b_n=2n-\frac{1}{3},
\end{equation}
 to the 1D surface brightness profile $I(R)$. 
The minimizations are done with a Levenberg-Marquardt algorithm, in a least squared sense.  The fits are
bounded by adding a square penalty function for parameters outside the
specified range. The boundaries are: $\min[I(R)]
<I_{n,1D}<\max[I(R)]$, $1<R_{n,1D}< \max[R]$, $\frac{1}{2}<n_{1D}<50$.
The output parameters are $I_{n,1D}$, $R_{n,1D}$, $n_{1D}$.

The 2D fitting applies  Eq.~(\ref{eq:leisersic}) convolved with the PSF and with 
$R$ replaced by $R=\sqrt{x'^2 + y'^2/q_{2D}^2}$, where 
\begin{eqnarray}
x' = \ (x-x_{0}^{2\!D})\cos(PA^{2\!D}) - (y-y_{0}^{2\!D})\sin(PA^{2\!D}) \\
y' = -(x-x_{0}^{2\!D})\sin(PA^{2\!D}) - (y-y_{0}^{2\!D})\cos(PA^{2\!D}).
\end{eqnarray} 
Coordinates $x,y$ refers positions in the galaxy image.  The
two-dimensional S\'ersic function is fitted directly to the galaxy
image, except that pixels outside the galaxy, as defined by the
Petrosian Region, flagged stars and central circular region of 1 PSF
FWHM are masked. The algorithm is the same as for 1D fitting, with the
following boundaries: the center $(x_0,y_0)_{2D}$ cannot vary more
than 15\% compared to $(x_0,y_0)_{\rm peak}$; $I_{n,2D}$ must be
within image pixel values range; $R_{n,2D}$ cannot be greater than the
image half-diagonal; $\frac{1}{2} <n_{2D}<20$,
\mbox{$\frac{1}{10}<q_{2D}=b/a<1$}. This setup has proven in
simulation and in real galaxies to be the most stable, converging for
most galaxies in the samples. The fit free parameters are
$\{x_0,y_0,PA, q,I_n, R_n, n\}_{2D}$.

Based on simulations of synthetic S\'ersic galaxies, we found that the
2D fitting is better than 1D at recovering "true" parameters from
images. Since the 2D is more unstable to initial parameters, we use
the 1D results as the initial guess for the 2D fit.

\subsection{Quality Flags}
\label{ssec:QF}

For reference, a series of conditions are evaluated and informative
\textbf{Quality Flags} (QF) are saved. They are not conclusive but may
indicate situations when the condition occur. For example, if for a
given object the $R_{n,2D}$ is of the order of the PSF, $n_{2D} \sim
0.5$ (a Gaussian) and $b/a\sim 1$, the object is probably a star, a
\mfmtk\ target selection error. Other QF indicate that the fitting
routine did not converge in situations of a crowded field. \rrbf{ For
  detecting crowded fields, we define the asymmetry $A_4$ as the
  distance between $\mathbf{r}_{\rm CoL} = (x_0,y_0)_{\rm CoL}$ and
  $\mathbf{r}_{\rm peak} = (x_0,y_0)_{\rm peak}$ in units of $R_p$, in
  percentage,
\[
A_4 = 100 \frac{\textbf{r}_{\rm CoL} - \textbf{r}_{\rm peak}}{R_p},
\]
which attains values greater than $\sim 10$ in crowded fields, and is
used to turn the QF64 on.}

 The QF are summarized in Table \ref{tab:QF}.

\begin{table}
	\begin{tabular}{lrrrr} \hline
		Flag  &   NAME       &   DEC  &  HEX   & CRITERIA \\ \hline          
		QF0   & normal      & 0  &   0x00  & no unusual situation \\
		QF1   & targetsize   & 1  &   0x01  &   {\tt psf\_fwhm}  $>R_p$  \\
		QF2   & targetisstar & 2  &   0x02 &   $R_{n,2D} \leq \mathtt{psf\_fwhm}$  and $n_{2D}\leq 0.55$ and  $b/a>0.8$ \\
		QF4   & fit1Derror   & 4  &   0x04 &   1D fitting routine did not converge \\
		QF8   & fit2Derror   & 8  &   0x08 &  2D fitting routine did not converge  \\ 
		QF16  & crowded1     & 16 &   0x10 &   more than 5\% of Petro region masked as stars  \\
		QF32  & crowded2     & 32 &   0x20 &   $R_{n,2D} > 2 R_p$\footnote{too many objects make this}  \\
		QF64  & crowded3     & 64 &   0x40 &   $A_4>10$  \\
	\hline
	\end{tabular} 
	\label{tab:QF}
	\caption{\mfmtk\ quality flags used that mark unusual situations  }
\end{table}

\subsection{Petrosian Quantities}
\label{sec:petrosian}
Petrosian (1976) defined a function $\eta(R)$ which is the ratio of
the mean intensity inside $R$ to the intensity at the isophote $R$
\begin{align}
\label{eq:petrosian_eta}
\eta(R) = \frac{\langle I \rangle (R)} {I(R)}.
\end{align}
The \textbf{Petrosian radius} is the distance from the galaxy center where the
fraction in Eq.(\ref{eq:petrosian_eta}) has some constant value 
\[
\eta(R_p) = \eta_0.
\]
Here we use $\eta_0=5$.  The virtue of $\eta$ is that both the
numerator and denominator have the same dependence with the distance,
hence $\eta$ is distance independent. The Petrosian radius is used as
a implicit scale length for each galaxy.

\section{Petrosian radius and S\'ersic index equivalence}
\label{app:Rp_and_n}

\begin{figure}
	\centering
    \includegraphics[width=.5\textwidth]{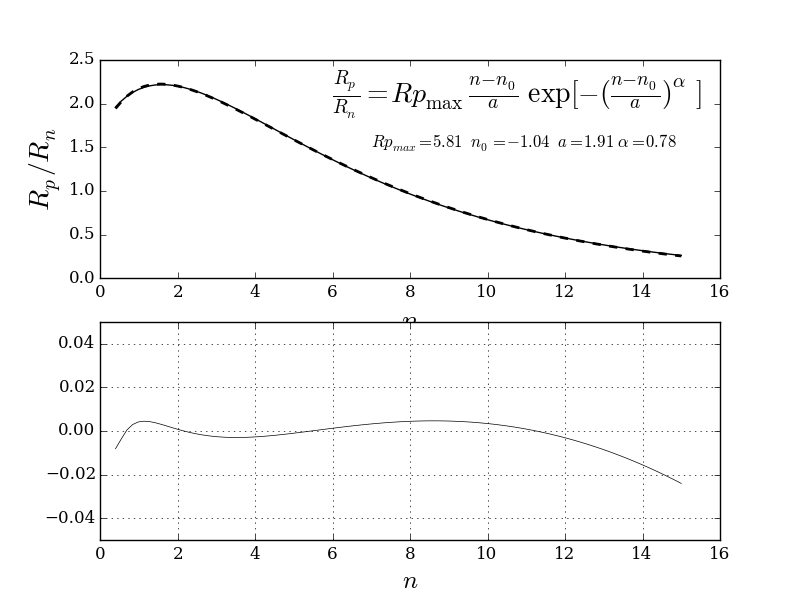}
	\caption{Petrosian radius as a function of $n$.  \textbf{Top:} numerical
		(dashed) and empirical function (continuous). \textbf{Bottom:} fractional
		error between the two. }
	\label{fig:petrosian_radius}
\end{figure}

The mean intensity within radius $R$ for a Sérsic model is the
integrated luminosity up to $R$ divided by the region area $\langle
I\rangle = L(R)/A$, $A=\pi R^2$, so for $x\equiv b_n(R/R_n)^{1/n}$, we
have (see for example \cite{Ciotti,graham})
\begin{align}
\langle I \rangle (R) =  2n \, I_n\, e^b \; \frac{\gamma(2n,x)}{x^{2n}}.
\end{align}
We then have for the Petrosian function Eq. \ref{eq:petrosian_eta}
\begin{align}
\eta(R) = \frac{2n \gamma(2n,x) }{x^{2n} e^{-x}}.
\end{align}
We have to solve 
\begin{align}
\frac{2n \gamma(2n,x_p) }{x_p^{2n} e^{-x_p}} = \eta_0 \quad \text{with} \quad x_p=x(R_p),
\end{align}
to obtain $R_p$ as a function of $n$. This equation is transcendental and can
only be solved for $R_p$ numerically. However, for practical purposes, we can
write an empirical Petrosian Radius function
\begin{align}
R_p(n) = R_n \; R_p^{\rm max} \ \frac{n-n_0}{a} \exp\left[ - \left(
\frac{n-n_0}{a} \right)^\alpha \right]
\end{align}
whose parameters $R_{\rm max}=5.8$, $n_0=-1.11$, $a=2.04$ and $\alpha=0.8$
provide a fit better than 1\% over the range $0.3<n<15$, as shown in Figure~\ref{fig:petrosian_radius}.

\section{Concentration and S\'ersic index equivalence}
\label{app:C_and_n}

\begin{figure}
	\centering
	\includegraphics[width=.5\textwidth]{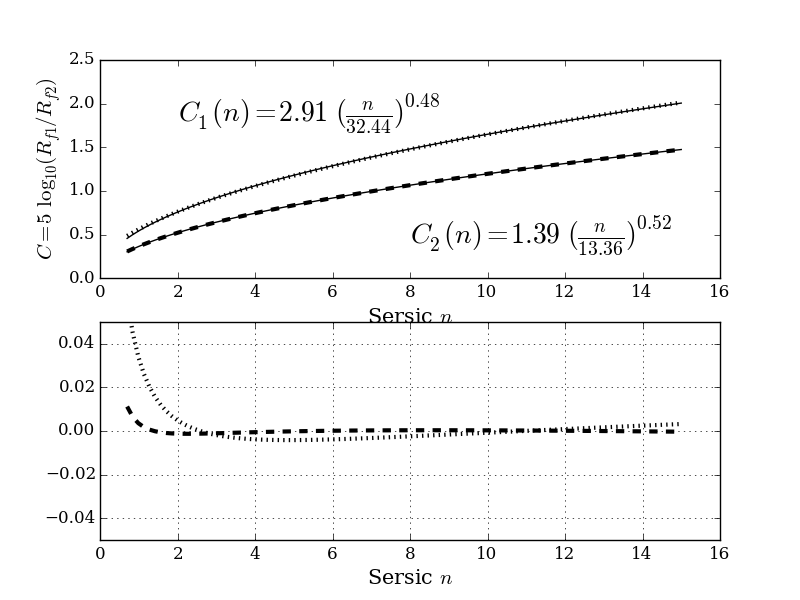}
	\caption{Concentration as a function of $n$. \textbf{Top:}
          numerical (dotted for $C_1$, dashed for $C_2$) and empirical
          function (continuous). \textbf{Bottom:} fractional error
          between the two.}
	\label{fig:concentration_sersic}
\end{figure}

In the case of Sérsic law, the integrated luminosity within radius $R$ \citep{Ciotti} is  
\begin{equation}
	\label{eq:LdeR}
	L(R) = 2 \pi n \, I_n R_n^2 \, \frac{e^b}{b^{2n}} \; \gamma(2n,x)
\end{equation}
with $x\equiv b_n (R/R_n)^{1/n}$. Hence the total Luminosity $L_T = L(R\to\infty)$ is
\begin{equation}
	\label{eq:LT}
	L_T = 2 \pi n \, I_n R_n^2 \, \frac{e^b}{b^{2n}} \; \Gamma(2n).
\end{equation}
From Eqs. \ref{eq:LdeR} and \ref{eq:LT} we have the equation for the
$R_f$ which attains some fraction $f$ of the total luminosity
\begin{align}
\gamma(2n,x_f) = f\ \Gamma(2n) \qquad \textrm{with} \qquad  x_f = x(R=R_f)
\end{align}
or for both $R_{f1}$ and $R_{f2}$
\begin{align}
\label{eq:xf}
\frac{\gamma(2n,x_{f1})}{\gamma(2n,x_{f2})} = \frac{f_1}{f_2}
\end{align}
Equation~(\ref{eq:xf}) cannot be solved analytically (except for
$n=1/2$) and the solution must be found
numerically. Figure~\ref{fig:concentration_sersic} shows the numerical
solution for $1/2<n<15$.

Again, we can write an  empirical function 
\begin{align}
C(n) = C' \left(\frac{n}{n'}\right)^\beta
\end{align}
which approximates the solution in the specified range with an error
smaller than 2\% for $C_1$ in the range $1<n<15$, with $C'=2.91$,
$n'=32.44$ and $\beta=0.48$

\section{Histogram of Morphometric Parameters for Databases}
\label{app:histogramas}

\begin{figure}
\centering
\includegraphics[width=.7\textwidth]{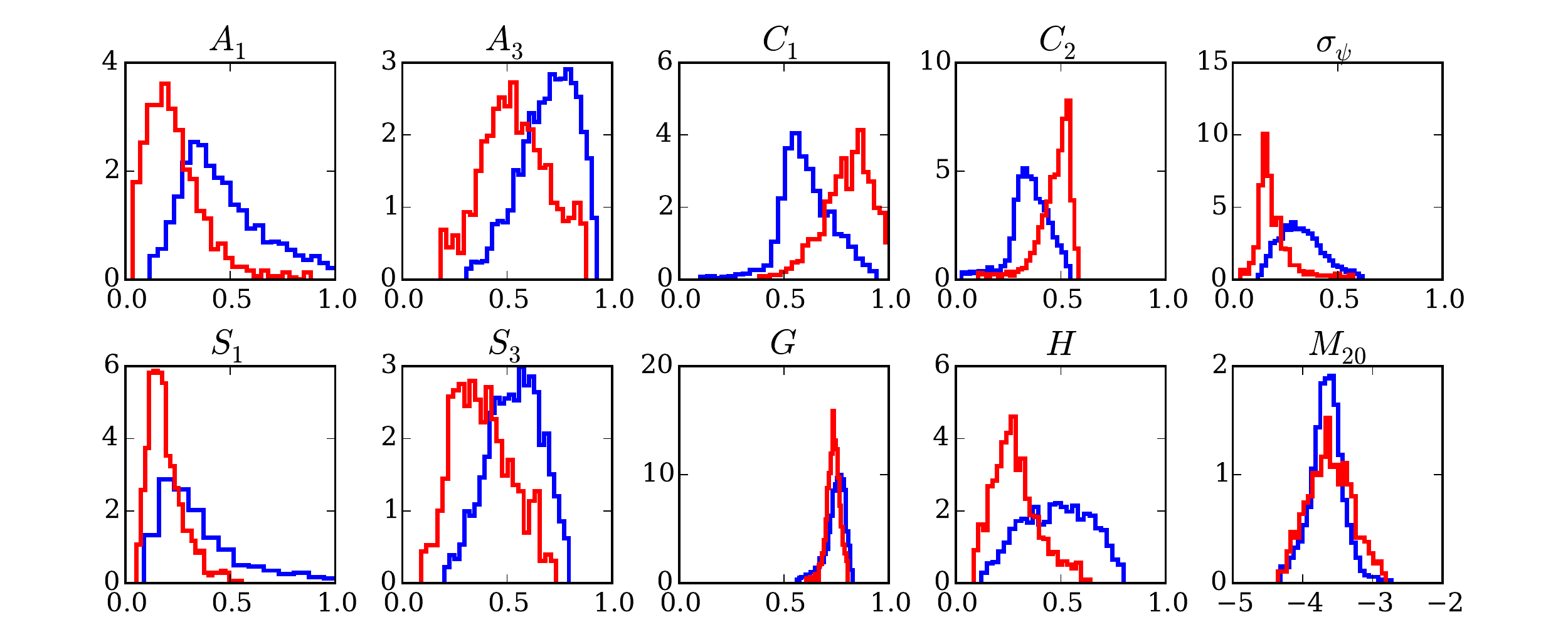}
\caption{Distribution of feature values among morphometric classes for the EFIGI database.}
\label{fig:EFIGI_ima_r_v47_T_distrofit}
\end{figure}
\begin{figure}
\centering
\includegraphics[width=.7\textwidth]{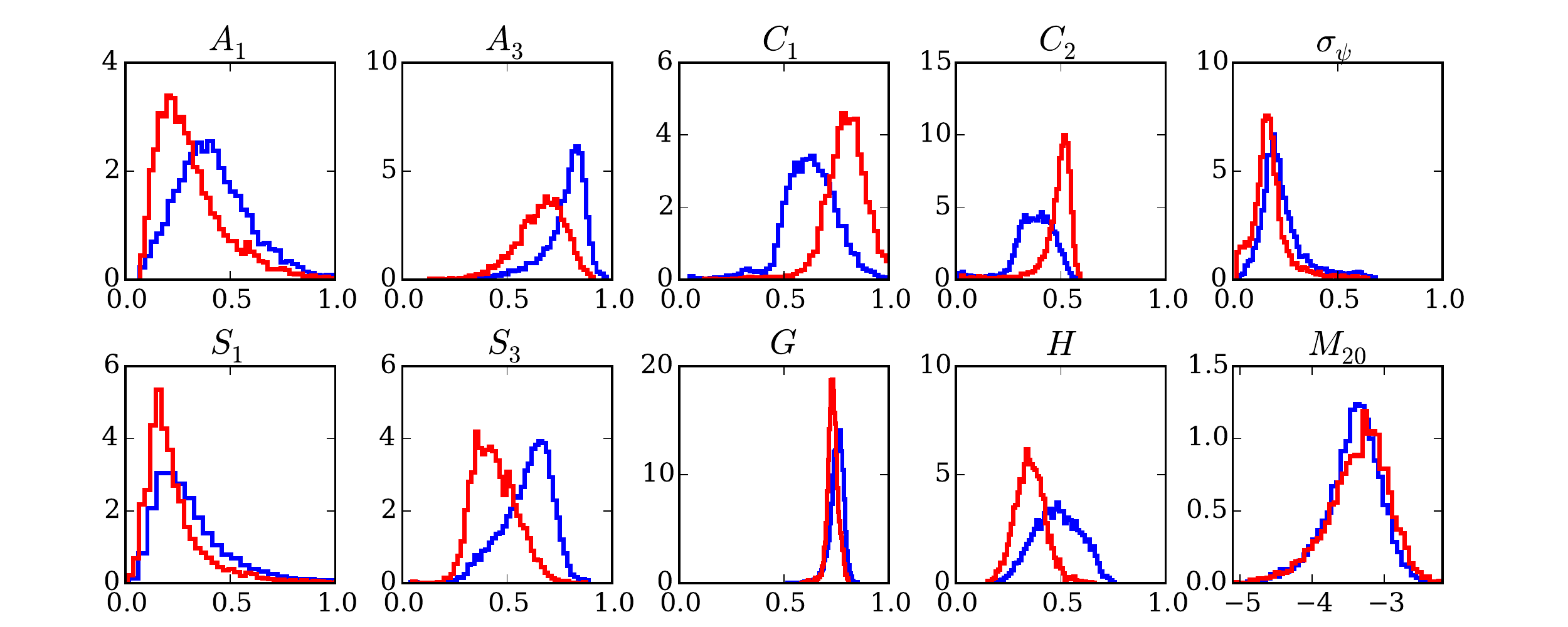}
\caption{Distribution of feature values among morphometric classes for the NA database.}
\label{fig:Abraham_stamps_v47_NAtable2_distrofit}
\end{figure}
\begin{figure}
\centering
\includegraphics[width=.7\textwidth]{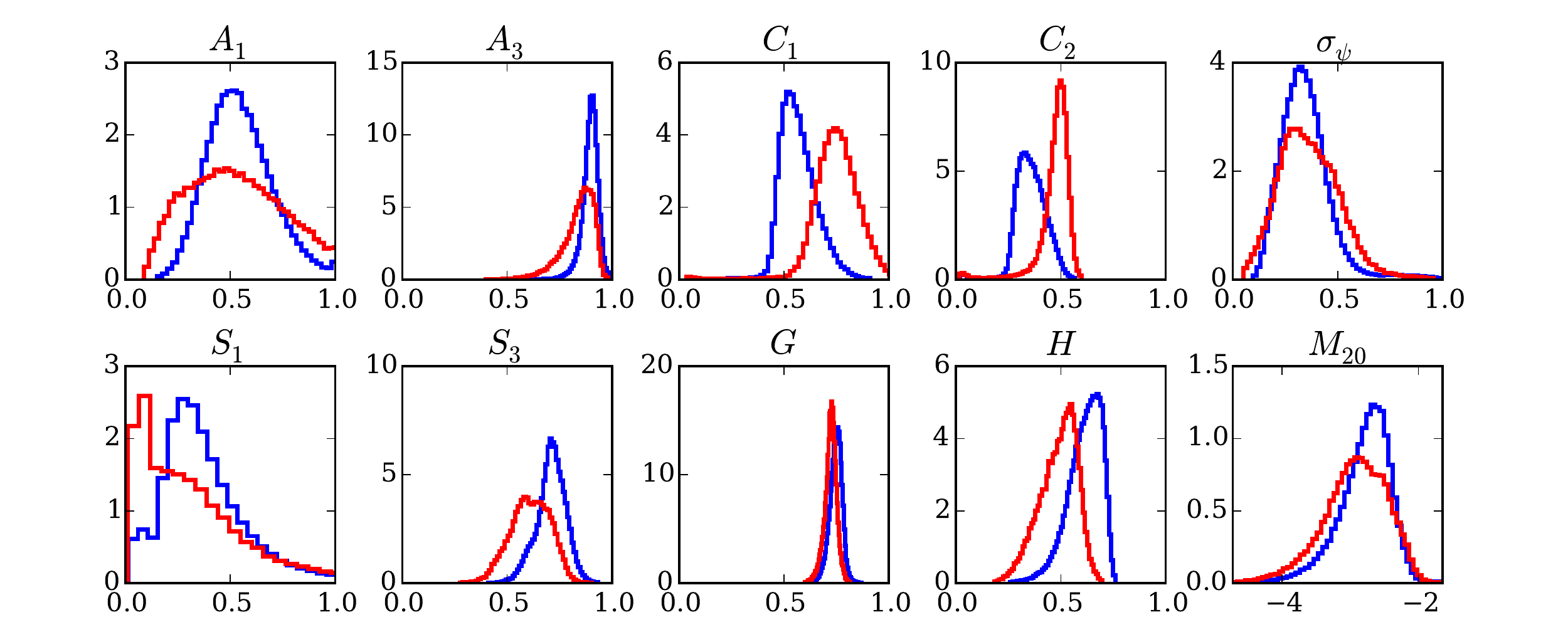}
\caption{Distribution of feature values among morphometric classes for the LEGACY complete sample.}
\label{fig:LEGACY_run01_zoo_distrofit}
\end{figure}
\begin{figure}
	\centering
	\includegraphics[width=.7\textwidth]{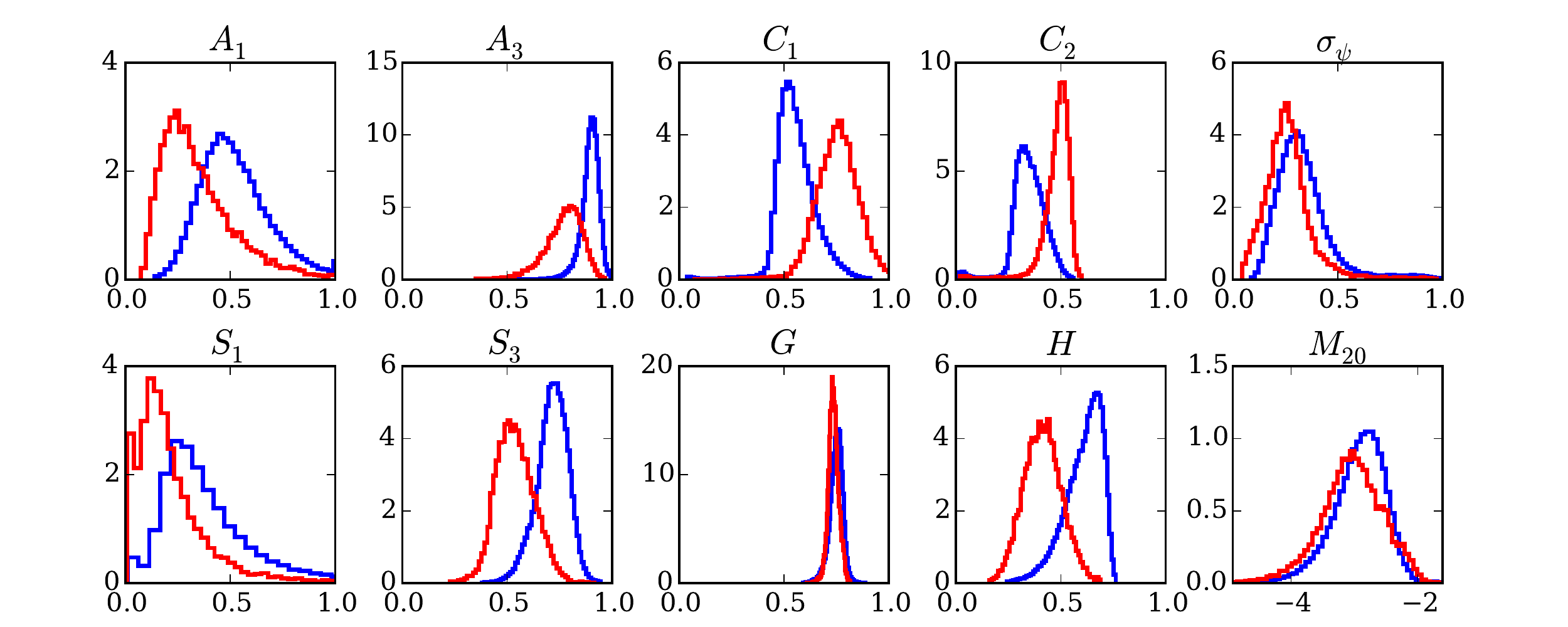}
	\caption{Distribution of feature values among morphometric classes for the LEGACY--$zr$ sample.}
	\label{fig:LEGACY_run01_zoo_z01r1778_distrofit}
\end{figure}


\end{document}